\documentclass{article} 
\usepackage{iclr2026_conference,times}

\usepackage{amsmath,amsfonts,bm}

\def\eqref#1{equation~\ref{#1}}

\def\1{\bm{1}}

\DeclareMathAlphabet{\mathsfit}{\encodingdefault}{\sfdefault}{m}{sl}
\SetMathAlphabet{\mathsfit}{bold}{\encodingdefault}{\sfdefault}{bx}{n}

\usepackage{amsmath}
\usepackage{wrapfig}
\usepackage{inconsolata}
\usepackage{microtype}
\usepackage{times}
\usepackage{latexsym}
\usepackage{subcaption}
\usepackage[colorlinks=true, allcolors=blue]{hyperref}
\usepackage{url}
\usepackage{graphicx}
\usepackage{arydshln}
\usepackage{booktabs} 
\usepackage{multirow} 
\usepackage{makecell} 
\usepackage{tabularx}
\usepackage{array}  
\usepackage{adjustbox}
\usepackage{tcolorbox}
\usepackage{xcolor}
\usepackage{colortbl}  
\definecolor{red}{rgb}{0.99, 0.02, 0.02}
\definecolor{mgreen}{RGB}{0,200,0}

\definecolor{demphcolor}{gray}{.45}
\newcommand{\demph}[1]{\textcolor{demphcolor}{#1}}

\title{\textsc{SweRank}: Software Issue Localization \\ with Code Ranking}

\author{Revanth Gangi Reddy\thanks{Equal Contribution. Work done during Revanth's internship at Salesforce AI Research.}$\hspace{0.4em}^{1,2}$\hspace{0.5em}Tarun Suresh\footnotemark[1]$\hspace{0.4em}^1$\hspace{0.5em}JaeHyeok Doo\footnotemark[1]$\hspace{0.4em}^3$\hspace{0.5em}Ye Liu$^2$\hspace{0.5em}Xuan Phi Nguyen$^2$\\\hspace{3em}\textbf{Yingbo Zhou}$^2$\hspace{0.7em}\textbf{Semih Yavuz}$^2$ \hspace{0.7em}\textbf{Caiming Xiong}$^2$\hspace{0.7em}\textbf{Heng Ji}$^1$\hspace{0.7em}\textbf{Shafiq Joty}$^2$\\
$^1$ University of Illinois at Urbana-Champaign\hspace{1em}$^2$ Salesforce Research\hspace{1em}$^3$ KAIST AI}

\iclrfinalcopy 
\begin{document}

\maketitle

\begin{abstract}
Software issue localization, the task of identifying the precise code locations (files, classes, or functions) relevant to a natural language issue description (e.g., bug report, feature request), is a critical yet time-consuming aspect of software development. While recent LLM-based agentic approaches demonstrate promise, they often incur significant latency and cost due to complex multi-step reasoning and relying on closed-source LLMs. Alternatively, traditional code ranking models, typically optimized for query-to-code or code-to-code retrieval, struggle with the verbose and failure-descriptive nature of issue localization queries. To bridge this gap, we introduce \textsc{SweRank}\footnote{Code and models are available here: \url{https://github.com/SalesforceAIResearch/SweRank}}, an efficient and effective retrieve-and-rerank framework for software issue localization. To facilitate training, we construct \textsc{SweLoc}, a large-scale dataset curated from public GitHub repositories, featuring real-world issue descriptions paired with corresponding code modifications. Empirical results on SWE-Bench-Lite and LocBench show that \textsc{SweRank} achieves state-of-the-art performance, outperforming both prior ranking models and costly agent-based systems using closed-source LLMs like Claude-3.5. Further, we demonstrate \textsc{SweLoc}'s utility in enhancing various existing retriever and reranker models for issue localization, establishing the dataset as a valuable resource for the community.
\end{abstract}

\section{Introduction}
\label{sec:introduction}

The scale and complexity of modern software systems continue to grow exponentially, with a significant portion of development effort dedicated to identifying and resolving software issues. This has fueled growth in automated software issue fixing~\citep{cognition2024devin}, with recent LLM-based patch generation~\citep{yang2024swe, aider2024swe} solving real-world issues on benchmarks such as SWE-Bench~\citep{jimenez2023swe}, and commercial copilots integrating ``one‑click'' quick‑fix suggestions directly into IDEs~\citep{microsoft2023copilot, cursor2025, windsurf2025}.
Central to the process of fixing software issues 
is the task of \textbf{issue localization}: accurately identifying \textit{where} in the codebase the necessary changes should be made. This involves pinpointing the specific files, classes, or functions relevant to a given issue description, typically provided in natural language (e.g., a bug report). 
Effective localization is critical; without correctly identifying the relevant code segments, any subsequent attempt at automated repair is likely to fail or, worse, introduce new faults.

Given the importance of localization, recent work treats it as an agentic reasoning problem~\citep{yao2023react} and has investigated the use of sophisticated LLM-based agents~\citep{yang2024sweagent, yu2025orcaloca, chen2025locagent} that issue commands such as `read-file', `grep' and `traverse-graph' to iteratively explore codebases, navigate file structures, search for code patterns, and analyze dependencies. While powerful, these agent-based compound  systems often involve multiple rounds of interaction ($\approx$7–10 on average) with large models and complex reasoning processes, which can incur considerable API costs ($\approx$\$0.66 per example with Claude‑3.5) at high latency.  Moreover, agent traces are brittle: they rely on temperature sampling and require complex tool orchestration.

An alternative, more efficient strategy is to frame issue localization as an information retrieval problem, specifically using code ranking models~\citep{wang2021codet5, zhang2024code, suresh2024cornstack}. Such models can directly rank candidate code snippets (e.g., functions or files) based on their relevance to a given natural language  query, and quickly score and sort potential locations within a large codebase. However, prior code ranking models are still inferior in performance as they have predominantly been optimized for tasks distinct from issue localization. These typically include query-to-code retrieval~\citep{li2024codesearchnet}, which aims to find code implementing a described functionality, and code-to-code retrieval~\citep{Wang-FSE23,li2024coir}, focused on identifying semantically similar code fragments. 
The task of issue localization presents unique characteristics; input queries (issue descriptions) are often substantially more verbose than typical NL-to-code queries\footnote{\hspace{0.2em}460 tokens in SWE-Bench~\citep{jimenez2023swe} issues vs 12 tokens in CSN~\citep{li2024codesearchnet} queries.} and, more crucially, issues tend to describe observed erroneous behavior or system failures rather than specifying desired functionality. This fundamental difference in query nature and intent suggests that models trained on conventional code retrieval data~\citep{husain2019codesearchnet, suresh2024cornstack} may not be optimally suited for issue localization.
\begin{wrapfigure}{r}{0.44\linewidth}
    \vspace{0.5em}
    \centering
    \includegraphics[width=1.0\linewidth]{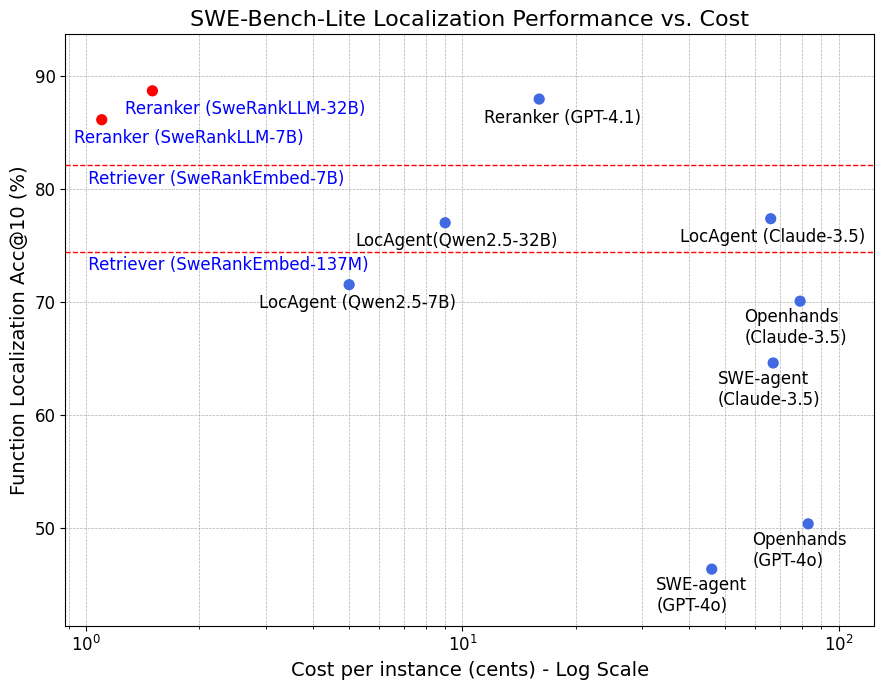}
    \vspace{-1.6em}
    \caption{\small Comparison of localization performance versus cost per instance on SWE-Bench-Lite. Our proposed \textsc{SweRankEmbed} retriever and \textsc{SweRankLLM} reranker models achieve superior accuracy at a significantly lower cost compared to agent-based localization methods.}
    \label{fig:perf_vs_cost}
    \vspace{-1.5em}
\end{wrapfigure}

To bridge this gap, we introduce \textsc{SweRank}, a code ranking framework trained specifically for software issue localization. \textsc{SweRank} employs a standard yet effective retrieve-and-rerank architecture, comprising two core components: (1) \textsc{SweRankEmbed}, a bi-encoder embedding model serving as the code retriever; and (2) \textsc{SweRankLLM}, an instruction-tuned LLM serving as a code reranker. To train \textsc{SweRank}, we construct \textsc{SweLoc}, a new large-scale issue localization dataset curated from public Github repositories, providing realistic training examples. \textsc{SweRankEmbed} is trained using a contrastive objective, where the issue descriptions serve as queries, the known localized functions act as positive examples, and carefully mined code snippets from the same repository function as hard negatives. Subsequently, \textsc{SweRankLLM} is trained as a list-wise reranker~\citep{reddy2024first}; it takes as input the issue description alongside the top-$K$ candidates retrieved by \textsc{SweRankEmbed} and predicts an improved ranking permutation, thereby enhancing the final localization.

Empirical results demonstrate that \textsc{SweRank} achieves state-of-the-art performance for file, module and function-level localization on Swe-Bench-Lite~\citep{jimenez2023swe} and LocBench~\citep{chen2025locagent}. Further, we show that \textsc{SweRank}, built on open-source models, has a considerably better performance to cost ratio compared to agent-based approaches that employ closed-source LLMs like Claude-3.5~\citep{anthropic2023claude}, as illustrated in Figure~\ref{fig:perf_vs_cost}. Finally, we demonstrate the effectiveness of our \textsc{SweLoc} data by showing that it consistently improves localization performance when used for finetuning a variety of text and code-pretrained retriever and reranker models. 


\section{Related Work}
\label{sec:related_work}

\subsection{Software Issue Localization}

Software issue localization or Fault Localization (FL) aims to identify the specific code locations responsible for reported bugs. Traditional fault localization methods~\citep{wong2016survey} can be grouped into spectrum‐based and program‐analysis approaches. Spectrum‐based fault localization (SFL)~\citep{de2016spectrum, amario2024understanding} statistically associates test outcomes with executed code elements to rank statements or functions by their `suspiciousness' based on passing and failing test coverage. Complementary static and dynamic analyses exploit program structure--through call‐graph traversal~\citep{adhiselvam2015enhanced}, dependency analysis~\citep{ELSAKA201779}, or program slicing~\citep{soremekun2021locating}--to constrain the search space of potential bug locations. 
While these methods provide a statistical basis for finding faults, they require precise program models and cannot leverage the rich natural language context in bug reports.

Modern approaches instead use LLM-based agent frameworks that treat bug localization as a planning and searching problem. AgentFL~\citep{qin2024agentfl} incorporates a multi-agent system with a three step procedure involving interpreting the bug context, traversing the codebase and verifying the suspected fault. OpenHands~\citep{wang2024openhands} and SWE-Agent~\citep{yang2024sweagent} use bash commands or custom interfaces to navigate repositories and access files. Other agentic systems combine IR with tool use: MoatlessTools~\citep{orwall2023moatless} integrates a semantic code search engine into an agent’s loop to guide it to relevant files. More recently, LocAgent~\citep{chen2025locagent} constructs a graph of the codebase for an LLM agent to do multi-hop reasoning over code dependencies. While these agent-driven approaches have achieved impressive results, they incur substantial costs and have high latency. Agent-based methods must orchestrate multiple steps of reasoning and tool use, which makes them brittle; a single failure in the chain (e.g., a misleading intermediate query or an incomplete code observation) can derail the entire localization process.
\textsc{SweRank} instead formulates issue localization as a single-shot ranking problem, which is highly efficient and cost-effective. 

\subsection{Code Ranking}

Transformer-based code ranking models~\citep{wang2023codet5opencodelarge, zhang2024code, günther2023jinaembeddingsnovelset, suresh2024cornstack} have set state-of-the-art on a variety of code retrieval tasks~\citep{li2024codesearchnet, li2024coir} by learning joint embeddings of text and code.~\citet{wang2023codet5opencodelarge} and~\citet{zhang2024code} learn improved code representations by incorporating a mix of training objectives, such as span denoising, text-code matching and causal LM pretraining, over large-scale code corpora such as CodeSearchNet~\citep{husain2019codesearchnet} and The Stack~\citep{kocetkovstack}.~\citet{suresh2024cornstack} improve the contrastive training process between function snippets and associated doc-strings with better consistency filtering and harder negative mining.~\citet{liu2024codexembed} incorporate multi-task contrastive data that includes code contest generation~\citep{billah2024large}, code summarization~\citep{sontakke2022code}, code completion~\citep{liurepobench}, code translation~\citep{pan2024lost} and code agent conversation~\citep{jin2024teach}.
However, prior code ranking models rarely include error logs in their training data and are not optimized for issue localization, where queries are verbose bug reports rather than precise functionality requests. In contrast, \textsc{SweRank} is explicitly trained on \textsc{SweLoc}, a new automatically collected set of real-world issue reports paired with known buggy functions. By optimizing a bi-encoder retriever and a listwise LLM reranker on this task-specific data, \textsc{SweRank} directly aligns verbose bug descriptions with faulty code, thereby improving localization accuracy.
\section{\textsc{SweLoc}: Issue Localization Data}
\label{sec:dataset}

\begin{figure*}
    \centering
    \includegraphics[width=\textwidth]{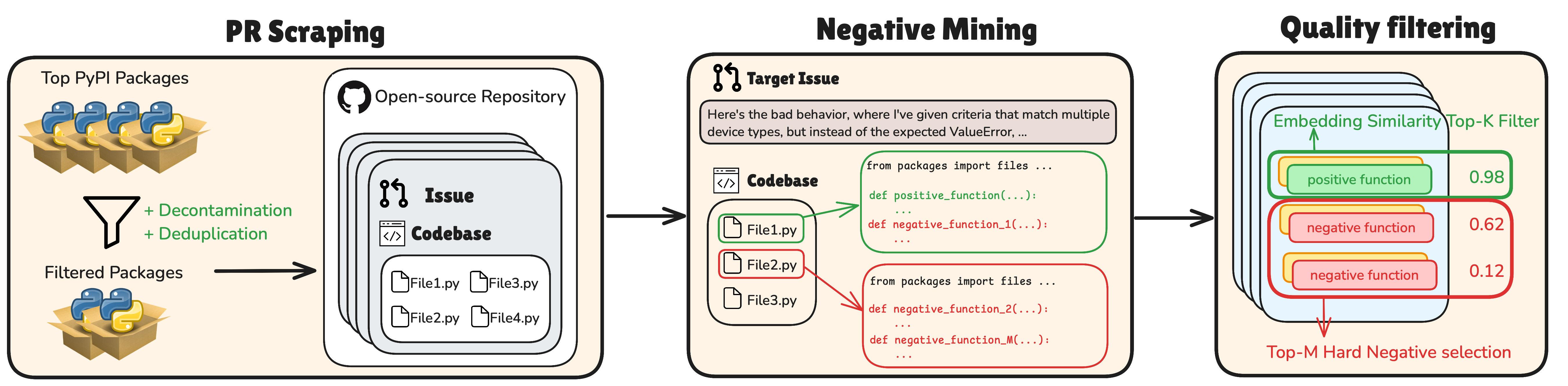}
    \vspace{-1em}
    \caption{\small Overview of \textsc{SweLoc} data construction pipeline, illustrating the three main stages.}
    \label{fig:sweloc_figure}
\end{figure*}

Existing code retrieval datasets~\citep{husain2019codesearchnet, suresh2024cornstack} are generally valuable for tasks like NL-to-code search which mainly requires functionality matching. However, they are sub-optimal for training models aimed at software issue localization. The nature of software issues--often detailed descriptions of failures rather than concise functional specifications--necessitates a dataset that accurately reflects this challenge of precisely identifying the problematic functions. To address this gap and provide a suitable training ground for our \textsc{SweRank} framework, we constructed \textsc{SweLoc}, a novel large-scale dataset specifically curated for the task of localizing code snippets  relevant to software issues. \textsc{SweLoc} is derived from real-world software development activities captured in public GitHub repositories. Our methodology comprises three main phases: (1) identifying and filtering relevant pull requests (PRs) from popular Python repositories (\S{\ref{sec:identifying_prs}}), (2) processing these PRs to extract issue descriptions paired with their corresponding code modifications (\S{\ref{sec:localization_preprocessing}}), and (3) applying consistency filtering and hard-negative mining to enhance the quality of training instances (\S{\ref{sec:minefilter}}). An overview of this process is shown in Figure~\ref{fig:sweloc_figure}.


\subsection{Identifying Relevant PRs}
\label{sec:identifying_prs}
Our data collection involves selecting the repositories associated with the top 11,000 PyPI packages on GitHub. To ensure repository quality and relevance to our task, we apply several filtering criteria. Repositories are required to contain at least 80\% Python code. To prevent data leakage and overlap with existing benchmarks, we exclude repositories already present in SWE-Bench~\citep{jimenez2023swe} and LocBench~\citep{chen2025locagent}. Finally, we perform deduplication based on source code overlap to remove near-identical repositories. This process results in a curated set of 3387 repositories.

\begin{figure*}
    \centering
    \begin{subfigure}[c]{0.44\linewidth}
        \centering
     \includegraphics[width=1\linewidth]{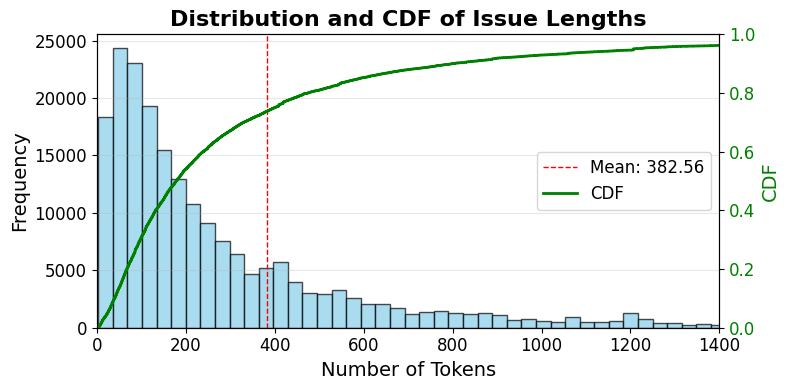}
     \label{fig:query_lens}
     \vspace{-0.75em}
    \end{subfigure}
    \hfill
    \begin{subfigure}[c]{0.54\linewidth}
        \centering
     \includegraphics[width=1\linewidth]{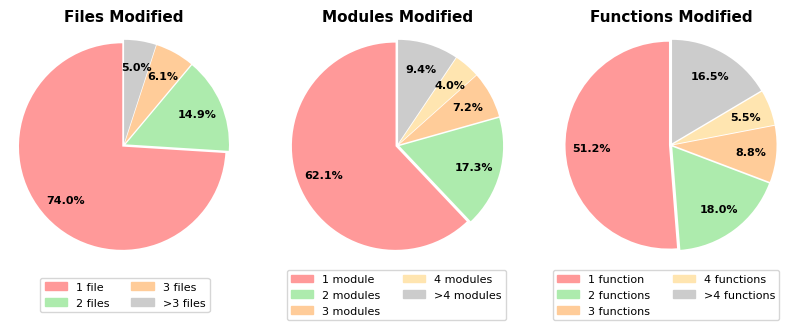}
     \label{fig:num_modified}
     \vspace{-0.75em}
    \end{subfigure}
    \vspace{-1em}
    \caption{\small (Left) Distribution of query lengths in the \textsc{SweLoc} dataset. The red dashed line indicates a mean query length of 382.56 tokens, underscoring the detailed nature typical of issue reports.
    (Right) Distribution of the number of (a) files, (b) modules, and (c) functions modified per GitHub issue. This highlights that while many localizations are concentrated, a significant number span multiple code units, particularly at finer granularities.}
    \label{fig:data_dist}
\end{figure*}

Following the SWE-Bench methodology, we identify pull requests (PRs) within these repositories that (1) resolve a linked GitHub issue and (2) include modifications to test files, indicating the issue resolution was verified. For each such PR, we collect the issue description and the codebase snapshot at the PR's base commit.
This procedure results in 67,341 initial $(\text{PR}, \text{codebase})$ pairs. Figure~\ref{fig:data_dist} provides further details on the dataset's composition, including query and repository edit distributions.

\subsection{Localization Processing}
\label{sec:localization_preprocessing}

Using the collected $(\text{PR}, \text{codebase})$ pairs, we create contrastive training data in the form of $\langle$query, positive, negatives$\rangle$ tuples. For each tuple, the issue description serves as the query. Each function modified within the PR is designated as a positive example, corresponding to a distinct training instance. Thus, a PR modifying $N$ functions yields $N$ training instances. The negatives for each instance come from the unmodified functions within the corresponding codebase. This initial set of instances are further refined via consistency filtering and hard-negative mining, as described next.

\subsection{Consistency Filtering and Hard Negatives}
\label{sec:minefilter}

The quality of $\langle$query, positive, negatives$\rangle$ tuples used for training significantly impacts the ranking model performance~\citep{suresh2024cornstack}. Effective contrastive learning requires relevant positives and challenging negatives (semantically similar to the positive but irrelevant to the query). However, issue descriptions in open-source repositories can be vague, leading to noisy signals for relevance between the issue descriptions and associated code modifications when directly used for training.

To mitigate this, we employ filtering and mining techniques following recent work~\citep{gunther2023jina, suresh2024cornstack}. First, we apply top-$K$ consistency filtering~\citep{suresh2024cornstack} to retain only instances where the positive code snippet is semantically close to the query relative to other code snippets in the repository. Formally, given an instance $i$ with issue description $t_i$, a positive function $c_i$, and the set of other unrelated functions $F_i$ in the repository, we use a pre-trained embedding model (\textsc{CodeRankEmbed}~\citep{suresh2024cornstack}) to compute similarities between $t_i$, $c_i$ and all functions in $F_i$. Instance $i$ is retained only if $c_i$ ranks within the top $K$ functions in $F_i$, based on similarity to $t_i$. We set $K=20$, with ablation studies in~\S{\ref{sec:data_quality_size}}.

Beyond filtering for relevance of positive pairs, incorporating challenging negatives is crucial for enabling the model to distinguish between semantically similar instances~\citep{moreira2024nv}. To this end, we employ a hard negative mining strategy that leverages the previously computed similarities to select a set of hard negatives $B_i = \{ c^-_j \}_{j=1}^{M}$ for each instance $i$. These negatives $c^-_j$ are chosen from $F_i$ such that they are among the top $M$ (=15) most similar functions to the query $t_i$.

\section{\textsc{SweRank} Methodology}
\label{sec:method}

In this section, we present our proposed ranking framework for software issue localization. \textsc{SweRank} adopts a two-stage retrieve-and-rerank approach with two key components: (1) \textsc{SweRankEmbed}, a bi-encoder retriever that efficiently narrows down candidate code snippets from large codebases; and (2) \textsc{SweRankLLM}, a listwise LLM reranker that refines these initial results for improved localization accuracy. Next, we elaborate on the architecture and training objectives for these components.

\subsection{\textsc{SweRankEmbed}}
\label{sec:swerankembed}
The retriever component, \textsc{SweRankEmbed}, utilizes a bi-encoder architecture~\citep{reimers-gurevych-2019-sentence} to generate dense vector representations for GitHub issues and code functions within a shared embedding space. Let $(t_i, c^{+}_i)$ represent a positive pair from the \textsc{SweLoc} dataset,  consisting of an issue $t_i$ and the corresponding code function modified $c^{+}_i$. The bi-encoder maps these to embeddings $(h_i, h^{+}_i)$, derived from the last hidden layer of the encoder. 
For a training batch of size $N$, let $H = \{h^+_i\}_{i=1}^N$ denote the set of positive code embeddings. Let ${H}_{B} = \bigcup_{i=1}^{n}\{h^-_{ij}\}_{j=1}^M$ be the set of embeddings for the $M$ hard negatives mined for each issue $t_i$ in the batch (as described in \S{\ref{sec:minefilter}}).

\textsc{SweRankEmbed} is trained using an InfoNCE contrastive loss~\citep{oord2018representation}. The objective encourages the embedding $h_i$ of an issue to have a higher similarity with its corresponding positive code embedding $h_i^+$, compared to its similarity with all other $h_k^+$ embeddings $(k\neq i)$ and all hard negative embeddings $h^-_{kj}$ within the batch. The loss for a single positive pair $(h_i,h_i^+)$ is:
\begin{equation}
\mathcal{L_{CL}} = 
- \log \left( 
    \frac{\exp (\mathbf{h}_i \cdot \mathbf{h^+}_{i})}
    { 
    \sum_{\mathbf{h}_k \in (\mathbf{{H}_{B}} \cup \mathbf{H})} \exp (\mathbf{h}_i \cdot \mathbf{h}_k)}
\right)
\end{equation}
The denominator sums over the positive embedding $h_i^+$ itself and $N(M+1)-1$ negative embeddings relative to $h_i$. 
During inference, candidate code functions for a given issue description are ranked based on the cosine similarity between their respective embeddings and the issue embedding.

\subsection{\textsc{SweRankLLM}}
\label{sec:swerankllm}

For the reranking stage, we employ \textsc{SweRankLLM}, an instruction-tuned LLM for reranking. \textsc{SweRankLLM} adopts a listwise ranking approach~\citep{pradeep2023rankzephyr}, which offers better performance than pointwise methods by considering the relative relevance of candidates. Typically, listwise LLM rerankers are trained to process an input consisting of the query and a set of candidate documents, each associated with a unique identifier. The model's training objective is then to generate the full sequence of identifiers, ordered from most to least relevant according to the ground-truth ranking.
However, since \textsc{SweLoc} does not provide a ground-truth ranking among the negative functions for the issue $t_i$, generating a complete target permutation for training is not feasible.

To adapt listwise reranking training to our setting where only the positive is known, we modify the training objective. Formally, let $\mathcal{D} := \{d_i\}_{i=1}^{|\mathcal{D}|}$ be a training dataset of triplets, where each sample $d_i := (t_i, c_i^{+}, \{c_{i,j}^{-}\}_{j=1}^{M})$ includes a GitHub issue $t_i$, a relevant positive code $c_i^{+}$, and a set of $M$ irrelevant negative codes $\{c_{i,j}^{-}\}_{j=1}^{M}$. We first assign a unique numerical identifier from 1 to $M$+1) to each function in the set $c_i^{+}\cup\{c_{i,j}^{-}\}_{j=1}^{M}$. Let $I_i^+$ be the identifier assigned to the positive function $c_i^+$. Instead of training the model to predict the full ranked list of identifiers, we train it to correctly generate the identifier corresponding to the single positive function, $I_i^+$. Thereby, the training objective for a given sample $d_i$ is thus simplified to maximizing the likelihood of the first generated (i.e. top-ranked) identifier:
\begin{equation}
\mathcal{L}_{LM} = -\log(P_{\theta}(I_i^+|x))
\end{equation}
where $x$ is the input prompt constructed from the issue $t_i$ and the set of candidate functions $c_i^{+}\cup\{c_{i,j}^{-}\}_{j=1}^{M}$ along with their assigned identifiers, and $P_\theta$ represents the listwise LLM reranker.

During training, we omit the end-of-sequence token after predicting $I_i^+$ to retain the model's capability to generate full ranked lists for inference, as required by the listwise format. As we show later in our experiments in \S{\ref{sec:choice_encoder}}, our approach enables finetuning any listwise reranker for the software issue localization task, without needing the full candidate ranking ordering for training supervision. 

 \begin{table}[t]
    \centering
    \renewcommand{\arraystretch}{1.2}
    \small
    \resizebox{1\textwidth}{!}{
    \begin{tabular}{l|ll ccc cc cc}
        \toprule
        \multirow{2}{*}{\textbf{Type}} & \multirow{2}{*}{\textbf{Method}} & \multirow{2}{*}{\textbf{Model}} & \multicolumn{3}{c}{\textbf{File} (\%)} & \multicolumn{2}{c}{\textbf{Module} (\%)} & \multicolumn{2}{c}{\textbf{Function} (\%)}\\
        \cmidrule(lr){4-6} \cmidrule(lr){7-8} \cmidrule(lr){9-10}
         & & & \textbf{Acc@1} & \textbf{Acc@3} & \textbf{Acc@5} & \textbf{Acc@5} & \textbf{Acc@10} & \textbf{Acc@5} & \textbf{Acc@10} \\
        \midrule
        \midrule
        
        \multirow{9}{*}{Agent} 
            & \multirow{2}{*}{\makecell[l]{MoatlessTools\\\cite{orwall2023moatless}}} 
                & \cellcolor{gray!10}\texttt{GPT-4o}   & \cellcolor{gray!10}73.36 & \cellcolor{gray!10}84.31 & \cellcolor{gray!10}85.04 & \cellcolor{gray!10}74.82 & \cellcolor{gray!10}76.28 & \cellcolor{gray!10}57.30 & \cellcolor{gray!10}59.49  \\
            & & \cellcolor{gray!10}\texttt{Claude-3.5} & \cellcolor{gray!10}72.63 & \cellcolor{gray!10}85.77 & \cellcolor{gray!10}86.13 & \cellcolor{gray!10}76.28 & \cellcolor{gray!10}76.28 & \cellcolor{gray!10}64.60 & \cellcolor{gray!10}64.96  \\
        \cmidrule(lr){2-10}
            & \multirow{2}{*}{\makecell[l]{SWE-agent\\\cite{yang2024sweagent}}}
                & \cellcolor{gray!10}\texttt{GPT-4o}   & \cellcolor{gray!10}57.30	&\cellcolor{gray!10}64.96	&\cellcolor{gray!10}68.98	&\cellcolor{gray!10}58.03	& \cellcolor{gray!10}58.03 &	\cellcolor{gray!10}45.99 & \cellcolor{gray!10}46.35 \\
            &                      & \cellcolor{gray!10}\texttt{Claude-3.5} & \cellcolor{gray!10}77.37 &	\cellcolor{gray!10}87.23&	\cellcolor{gray!10}90.15&	\cellcolor{gray!10}77.74&	\cellcolor{gray!10}78.10&		\cellcolor{gray!10}64.23&	\cellcolor{gray!10}64.60  \\
        \cmidrule(lr){2-10}
            & \multirow{2}{*}{\makecell[l]{Openhands\\\cite{wang2024openhands}}} 
                & \cellcolor{gray!10}\texttt{GPT-4o}   & \cellcolor{gray!10}60.95 & \cellcolor{gray!10}71.90 & \cellcolor{gray!10}73.72 & \cellcolor{gray!10}62.41 & \cellcolor{gray!10}63.87 & \cellcolor{gray!10}49.64 & \cellcolor{gray!10}50.36  \\
            & & \cellcolor{gray!10}\texttt{Claude-3.5} & \cellcolor{gray!10}76.28 & \cellcolor{gray!10}89.78 & \cellcolor{gray!10}90.15 & \cellcolor{gray!10}83.21 & \cellcolor{gray!10}83.58 & \cellcolor{gray!10}68.25 & \cellcolor{gray!10}70.07  \\
        \cmidrule(lr){2-10}
            & \multirow{3}{*}{\makecell[l]{LocAgent\\\cite{chen2025locagent}}}
            & \texttt{Qwen2.5-7B(ft)}      & 70.80 & 84.67 & 88.32 & 81.02 & 82.85 & 64.23 & 71.53  \\
            & & \texttt{Qwen2.5-32B(ft)} & 75.91 & 90.51 & 92.70 & 85.77 & 87.23 & 71.90 & 77.01  \\
            & & \cellcolor{gray!10}\texttt{Claude-3.5}   & \cellcolor{gray!10}77.74 & \cellcolor{gray!10}91.97 & \cellcolor{gray!10}94.16 & \cellcolor{gray!10}86.50 & \cellcolor{gray!10}87.59 & \cellcolor{gray!10}73.36 & \cellcolor{gray!10}77.37 \\
        \midrule
        \multirow{7}{*}{Retriever} 
            & \multicolumn{2}{l}{BM25~\citep{robertson1994okapi}} & 38.69    & 51.82 & 61.68  & 45.26    & 52.92    & 31.75	    & 36.86 \\
            & \multicolumn{2}{l}{Jina-Code-v2 (161M) ~\citep{günther2023jinaembeddingsnovelset}} & 43.43 & 71.17 & 80.29 & 63.50 & 72.63 & 42.34 & 52.19\\
            & \multicolumn{2}{l}{Codesage-large-v2 (1.3B)~\citep{zhang2024code}}  & 47.81 & 69.34 & 78.10 & 60.58 & 69.71 & 33.94 & 44.53\\
            & \multicolumn{2}{l}{CodeRankEmbed (137M)~\citep{suresh2024cornstack}} & 52.55 & 77.74 & 84.67 & 71.90 & 78.83 & 51.82 & 58.76 \\
            
            & \multicolumn{2}{l}{SFR-Embedding-2 (7B)~\citep{SFR-embedding-2}} & 58.03 & 80.29 & 	83.94 & 70.07 & 79.20 & 56.20 & 64.23 \\
            & \multicolumn{2}{l}{GTE-Qwen2-7B-Instruct (7B)~\citep{li2023towards}} & 65.33& 82.85& 89.78&76.28& 83.58&63.14&70.44\\
            \cdashline{2-10} 
            & \multicolumn{2}{l}{\textsc{\textbf{SweRankEmbed-Small}} (137M) (Ours)} & 66.42 & 86.50 & 90.88 & 79.56 & 85.04 & 63.14 & 74.45 \\
            & \multicolumn{2}{l}{\textsc{\textbf{SweRankEmbed-Large}} (7B) (Ours)} & \textcolor{blue}{72.63} & \textcolor{blue}{91.24} & \textcolor{blue}{94.16} & \textcolor{blue}{84.31} & \textcolor{blue}{89.78} & \textcolor{blue}{71.90} & \textcolor{blue}{82.12}	 \\
        \midrule
        \multirow{5}{*}{\makecell[l]{ + Reranker}} 
        & \multicolumn{2}{l}{CodeRankLLM (7B)~\citep{suresh2024cornstack}} & 72.99& 89.78 & 93.80 & 85.04 & 90.88 & 71.90 & 83.58 \\
       & \multicolumn{2}{l}{\cellcolor{gray!10}GPT-4.1} & \cellcolor{gray!10}82.12 &  \cellcolor{gray!10}95.62 & \cellcolor{gray!10}97.08 & \cellcolor{gray!10}93.07 & \cellcolor{gray!10}93.43 & \cellcolor{gray!10}81.75 & \cellcolor{gray!10}87.96 \\
       \cdashline{2-10} 
        & \multicolumn{2}{l}{\textbf{\textsc{SweRankLLM-Small (7B)}} (Ours)}  & 78.10 & 92.34 & 94.53 & 89.05 & 92.70 & 79.56 & 86.13 \\
         & \multicolumn{2}{l}{\textbf{\textsc{SweRankLLM-Large (32B)}} (Ours)}  & \textbf{83.21} &  \textbf{94.89} & \textbf{95.99}  & \textbf{90.88} & \textbf{93.43} & \textbf{81.39} & \textbf{88.69}\\
        \bottomrule
    \end{tabular}
    }
    \vspace{-0.5em}
    \caption{\small Performance (in \%) on SWE-Bench-Lite. The rerankers use \textsc{SweRankEmbed-Large} as the retriever. \colorbox{gray!10}{Gray} corresponds to results with closed-source models. Best retriever numbers are in \textcolor{blue}{blue}, while best overall numbers (except GPT-4.1) are in \textbf{bold}. }
    \label{tab:swebench_numbers}
    \vspace{-1em}
\end{table}
 \section{Experiments}
\label{sec:experiments}
The experiments compare \textsc{SweRank}'s performance against state-of-the-art agent-based localization methods, and other code ranking models (\S{\ref{sec:main_results}}). Furthermore, we investigate the impact of our \textsc{SweLoc} dataset, analyzing how its quality controls (such as consistency filtering) and size influence model performance (\S{\ref{sec:data_quality_size}}), and examining its generalizability by evaluating effectiveness in fine-tuning various pre-existing retriever and reranker models for the issue localization task (\S{\ref{sec:choice_encoder}}).

\subsection{Setup}
\label{sec:setup}
\paragraph{Model Training:} We train the \textsc{SweRank} models in two sizes: \textit{small} and \textit{large}. All models are finetuned using our \textsc{SweLoc} dataset. \textsc{SweRankEmbed-Small} is initialized with CodeRankEmbed~\citep{suresh2024cornstack}, a SOTA 137M code embedding model, while the large variant is initialized with GTE-Qwen2-7B-Instruct~\citep{li2023towards}, a 7B parameter text embedding model employing Qwen2-7B-Instruct as its encoder. The small version of \textsc{SweRankLLM} is initialized with \textsc{CodeRankLLM}~\citep{suresh2024cornstack}, a 7B parameter code-pretrained listwise reranker. The large version is initialized with Qwen-2.5-32B-Instruct that is pretrained using text listwise reranking data~\citep{pradeep2023rankzephyr}. More details in Appendix~\ref{sec:model_training}.

\paragraph{Baselines:} Our primary comparison is against prior agent-based localization methods. Specifically, we include OpenHands~\citep{wang2024openhands}, SWE-Agent~\citep{yang2024sweagent}, MoatlessTools~\citep{orwall2023moatless} and LocAgent~\citep{chen2025locagent}, the current SOTA agent-based approach. Notably, these methods predominantly use closed-source models, with LocAgent also finetuning open-source models for this task. For the retrieve-and-rerank framework, we compare \textsc{SweRankEmbed-Small} against BM25~\citep{robertson1994okapi} and several code embedding models of comparable size, including Jina-Code-v2~\citep{gunther2023jina}, Codesage-large-v2~\citep{zhang2024code}, and CodeRankEmbed~\citep{suresh2024cornstack}. For the 7B parameter embedding model comparison, we include GTE-Qwen2-7B-Instruct, which ranks third on the MTEB leaderboard~\citep{muennighoff2023mteb} at the time of evaluation. For the reranker comparison, we include \textsc{CodeRankLLM} and other closed source-models such as GPT-4.1. Due to the larger size of LocBench, comparisons on this benchmark are limited to a subset of the best-performing baselines.

\paragraph{Datasets \& Metrics:}We evaluate on SWE-Bench-Lite~\citep{jimenez2023swe} and LocBench~\citep{chen2025locagent}. Following~\citet{suresh2024cornstack}, we exclude examples from SWE-Bench-Lite where no existing functions were modified by the patch, resulting in 274 retained examples out of 300. While SWE-Bench-Lite primarily consists of bug reports and feature requests, LocBench ( 560 examples) also includes security and performance issues. Consistent with~\citet{chen2025locagent}, we measure localization performance at three granularities: file, module (class) and function, with Accuracy at $k$ (Acc@k) as the evaluation metric. This metric deems localization successful if all relevant code locations are correctly identified within the top-$k$ results. The relevance score for a specific file or module is determined by the maximum score of any function contained within that file or module.

\subsection{Localization Results}
\label{sec:main_results}
Table~\ref{tab:swebench_numbers} compares performance of different localization methods on the SWE-Bench-Lite benchmark. The results indicate that our \textsc{SweRank} models surpasses the performance of all evaluated agent-based methods. Furthermore, the \textsc{SweRankEmbed-Small} model, despite its relatively small size of 137M parameters, demonstrates highly competitive performance, outperforming prior 7B parameter embedding models. Notably, \textsc{SweRankEmbed-Large} achieves higher Acc@10 for function localization than LocAgent with Claude-3.5. Employing the \textsc{SweRankLLM} reranker subsequently enhances the retriever's output, establishing a new SOTA for localization performance on this benchmark across all granularities. Qualitative examples are provided in Appendix~\ref{sec:qualitative}.

\begin{wrapfigure}{r}{0.5\linewidth}
\vspace{-1.25em}
    \centering
    \includegraphics[width=\linewidth]{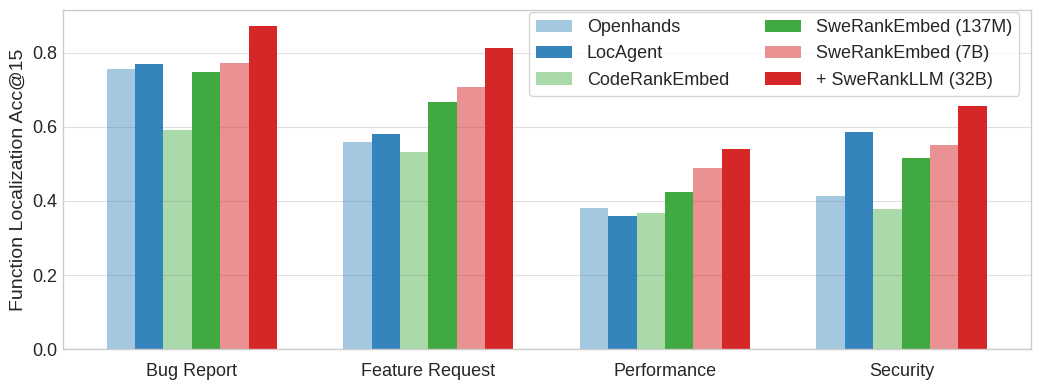}
    \vspace{-1.7em}
    \caption{\small Localization performance across different categories within LocBench. \textsc{SweRank} considerably outperforms Agent-based methods using Claude-3.5.}
    \vspace{-1.25em}
    \label{fig:category_wise}
\end{wrapfigure}
Table~\ref{tab:loc_res_new_bench} shows results on LocBench. A similar trend is observed, with the large variants of \textsc{SweRankEmbed} and \textsc{SweRankLLM} setting new SOTA performance. Figure~\ref{fig:category_wise} provides a detailed breakdown of localization accuracy across the four distinct difficulty categories within LocBench. Despite being primarily trained with bug reports in \textsc{SweLoc}, the \textsc{SweRank} models demonstrate impressive generalizability across other categories. 

\begin{table}[t]
    \centering
    \renewcommand{\arraystretch}{1.2}
    \resizebox{1.0\textwidth}{!}{
    \begin{tabular}{@{}l|c ccc cc cc@{}}
        \toprule
        \multirow{2}{*}{\textbf{Method}} & \multirow{2}{*}{\textbf{Loc Model}} & \multicolumn{2}{c}{\textbf{File} (\%)} & \multicolumn{2}{c}{\textbf{Module} (\%)} & \multicolumn{2}{c}{\textbf{Function} (\%)} \\
        \cmidrule(lr){3-4} \cmidrule(lr){5-6} \cmidrule(lr){7-8}
        & & Acc@5 & Acc@10 & Acc@10 & Acc@15 & Acc@10 & Acc@15 \\
        \midrule
        Agentless & \cellcolor{gray!10}\texttt{Claude-3.5} & \cellcolor{gray!10}67.50	&\cellcolor{gray!10}67.50	&\cellcolor{gray!10}53.39	&\cellcolor{gray!10}53.39	&\cellcolor{gray!10}42.68	&\cellcolor{gray!10}42.68 \\ 
        OpenHands & \cellcolor{gray!10}\texttt{Claude-3.5} & \cellcolor{gray!10}79.82	&\cellcolor{gray!10}80.00	&\cellcolor{gray!10}68.93	&\cellcolor{gray!10}69.11	&\cellcolor{gray!10}59.11	&\cellcolor{gray!10}59.29 \\ 
        SWE-agent & \cellcolor{gray!10}\texttt{Claude-3.5} & \cellcolor{gray!10}77.68	&\cellcolor{gray!10}77.68	&\cellcolor{gray!10}63.57	&\cellcolor{gray!10}63.75	&\cellcolor{gray!10}51.96	&\cellcolor{gray!10}51.96 \\ 
        \multirow{2}{*}{LocAgent} & \texttt{Qwen2.5-7B(ft)} & 78.57 &79.64 &63.04 &63.04 &51.43 &51.79 \\
     &\cellcolor{gray!10}\texttt{Claude-3.5} & \cellcolor{gray!10}83.39	&\cellcolor{gray!10}86.07	&\cellcolor{gray!10}70.89	&\cellcolor{gray!10}71.07	&\cellcolor{gray!10}59.29	&\cellcolor{gray!10}60.71 \\
        \midrule
        \multirow{4}{*}{Retriever} & CodeRankEmbed (137M) & 74.29 & 80.36 & 63.93 &	67.86 & 47.86 & 50.89 \\ 
        & GTE-Qwen2-7B-Instruct (7B) & 75.54 & 82.50 & 67.14 & 71.61 & 51.79 & 57.14	\\
        \cdashline{2-8} 
        & \textsc{\textbf{SweRankEmbed-Small} (137M)} & 80.36 & 84.82 & 71.43 & 75.00 & 	58.57 & 63.39	 \\
        & \textsc{\textbf{SweRankEmbed-Large} (7B)} & \textcolor{blue}{82.14} & \textcolor{blue}{86.96} & \textcolor{blue}{75.54} & \textcolor{blue}{78.93} & \textcolor{blue}{63.21} & \textcolor{blue}{67.32}	\\
        \midrule
        \multirow{5}{*}{+ Reranker}  & CodeRankLLM (7B) & 83.93 & 88.21 & 76.96 & 80.89 & 64.64 & 69.29	 \\
       & \cellcolor{gray!10}GPT-4.1 & \cellcolor{gray!10}85.89 & \cellcolor{gray!10}88.75 &\cellcolor{gray!10} 79.64 & \cellcolor{gray!10}82.50 & \cellcolor{gray!10}71.61 & \cellcolor{gray!10}74.64 \\
        \cdashline{2-8} 
        & \textsc{\textbf{SweRankLLM-Small}} (7B)& 85.54 & 88.39 & 79.11 & 82.14 & 69.46 & 74.46 \\
        & \textsc{\textbf{SweRankLLM-Large}} (32B)& \textbf{86.61} & \textbf{89.82} & \textbf{81.07} & \textbf{83.21} & \textbf{71.25} & \textbf{76.25}	\\
        \bottomrule
    \end{tabular}
    }
    \vspace{-0.5em}
    \caption{\small Performance (in \%) on LocBench. The rerankers use \textsc{SweRankEmbed-Large} as the retriever.  \colorbox{gray!10}{Gray} correspond to results with closed-source models. Best retriever model numbers are in \textcolor{blue}{blue}, while best overall numbers (except GPT-4.1) are in \textbf{bold}.}
    \label{tab:loc_res_new_bench}
    \vspace{-1em}
\end{table}

\subsection{Analysis}

Our analysis presented in this section aims to demonstrate the following key points: 1) the impact of \textsc{SweLoc} data quality and size on final model performance (\S{\ref{sec:data_quality_size}}); 2) the utility of \textsc{SweLoc} for finetuning various retriever and reranker models (\S{\ref{sec:choice_encoder}}; and 3) the cost-effectiveness of the proposed \textsc{SweRank} framework (\S{\ref{sec:cost_analysis}). Unless otherwise mentioned, the results are on SWE-Bench-Lite.


\subsubsection{Data Quality and Size}
\label{sec:data_quality_size}

Public GitHub repositories, as a source for contrastive data, often contain noisy instances. This study first examines the effectiveness of  consistency filtering (\S{\ref{sec:minefilter}}), specifically the influence of the positive-rank threshold, $K$. This parameter dictates the minimum rank of the instance's positive (relative to negatives, based on similarity with the issue description) for inclusion of the instance in the training set. Increasing $K$ relaxes the filtering, yielding more training instances but potentially introducing more noise. As shown in Figure~\ref{fig:k_filter}, finetuning \textsc{SweRankEmbed-Small} with \textsc{SweLoc} data filtered by different $K$ values reveals that optimal performance is achieved with a moderate $K$ (e.g., $K$=20), striking a balance between instance quality and dataset size. The absence of filtering ($K$=None) proves detrimental as performance drops after finetuning compared to pre-trained model.

Controlling for data quality (by fixing $K$=20), the impact of dataset size is investigated.  Figure~\ref{fig:data_size} illustrates that training with varying proportions of the filtered data yields considerable performance improvements, even with only 5\% of the data. Generally, larger dataset sizes correspond to further performance gains. These experiments underscore the significance of both data quality and quantity, demonstrating that merely increasing data volume without quality control can be detrimental. Appendix \S{\ref{sec:iterative_mining}} further examines the impact of negative hardness on model performance.

\begin{figure}[t]
    \begin{subfigure}[c]{0.54\linewidth}
        \includegraphics[width=1.0\linewidth]{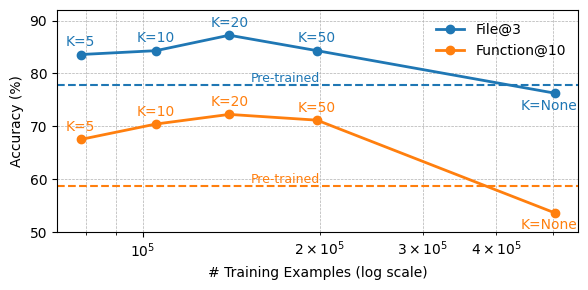}
        \vspace{-1.5em}
        \caption{\small While accuracy improves from doing consistency filtering, i.e. discarding instances where the positive's rank among negatives is $>$$K$, no filtering ($K$=\textit{None}) hurts performance.}
        \label{fig:k_filter}
    \end{subfigure}
\hfill
    \begin{subfigure}[c]{0.44\linewidth}
        \includegraphics[width=1.0\linewidth]{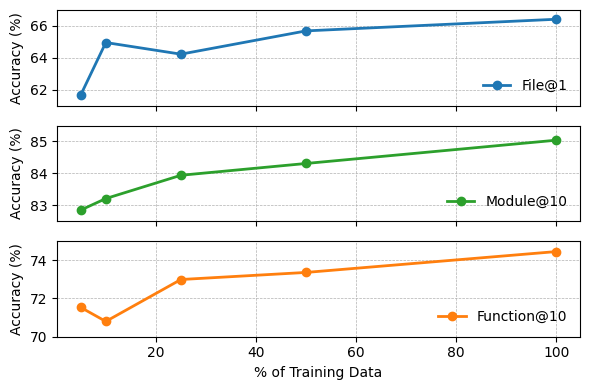}
        \vspace{-1.5em}
        \caption{\small All metrics show a general upward trend as the percentage of training data ($K$=20) increases.}
        \label{fig:data_size}
    \end{subfigure}
    \caption{\small Impact of (a) training data filtering and (b) data size on \textsc{SweRankEmbed-Small} performance.}
    \vspace{-0.5em}
\end{figure}

\subsubsection{Choice of Retriever and Reranker}
\label{sec:choice_encoder}

\begin{wraptable}{r}{0.43\textwidth}
\vspace{-1.1em}
\centering
\small
\renewcommand{\arraystretch}{1.35}
\setlength{\tabcolsep}{2.5pt}
\resizebox{\linewidth}{!}{
\begin{tabular}{@{}lccc@{}}
\toprule
\textbf{Base Retriever} & \textbf{Pretrain}  & \textbf{Func. Acc@10} (\%) \\
\midrule
CodeRankEmbed  & English+Code & 59.5$\rightarrow$\textbf{72.3} (\textcolor{mgreen}{+12.8})\\
Arctic-Embed & English  & 53.7$\rightarrow$71.9 (\textcolor{mgreen}{+17.4}) \\ 
Arctic-Embed-v2.0  & Multilingual & 62.0$\rightarrow$70.1 (\textcolor{mgreen}{+8.1}) \\
\bottomrule
\end{tabular}
}
\vspace{-0.6em}
\caption{\small Accuracy (Before$\rightarrow$After) from finetuning different retrievers with \textsc{SweLoc} data.}
\label{table:retriever_selection}
\vspace{-1em}
\end{wraptable}

Here, we demonstrate the effectiveness of \textsc{SweLoc} by showing improvements for a variety of retriever and reranker models from finetuning. First, the following embedding models, pre-trained on different data types, are finetuned for one epoch on \textsc{SweLoc}: Arctic-Embed~\citep{merrick2024arctic}, primarily pre-trained on English text retrieval data; CodeRankEmbed, pre-trained on 22 million NL-to-Code examples~\citep{suresh2024cornstack}; and Arctic-Embed-v2.0~\citep{yu2024arctic}, pre-trained on a mix of English and multilingual data. 
From Table~\ref{table:retriever_selection}, we see all models showing significant performance improvement from finetuning. Notably, models that initially performed weaker (e.g., Arctic-Embed) showed greater gains. This outcome validates that \textsc{SweLoc} can substantially improve the performance of \textit{any} embedding model for software issue localization.

\begin{wraptable}{r}{0.5\textwidth}
\centering
\small
\renewcommand{\arraystretch}{1.35}
\setlength{\tabcolsep}{3pt}
\resizebox{\linewidth}{!}{
\begin{tabular}{@{}lcc@{}}
\toprule
\textbf{Base LLM Reranker} & \textbf{Func. Acc@5} (\%)  & \textbf{Func. Acc@10} (\%)\\
\midrule
Qwen-2.5-Text (32B)& 77.0$\rightarrow$\textbf{81.4} (\textcolor{mgreen}{+4.4}) & 82.5 $\rightarrow$\textbf{86.1} (\textcolor{mgreen}{+3.6})\\
Qwen-2.5-Code (32B) & 76.3$\rightarrow$79.9 (\textcolor{mgreen}{+3.6}) & 81.8 $\rightarrow$84.7 (\textcolor{mgreen}{+2.9})\\
Qwen-2.5-Text (7B) & 75.2$\rightarrow$75.6 (\textcolor{mgreen}{+0.4})  & 81.4 $\rightarrow$82.5 (\textcolor{mgreen}{+1.1}) \\
Qwen-2.5-Code (7B) & 75.5$\rightarrow$75.9 (\textcolor{mgreen}{+0.4})  & 81.0 $\rightarrow$83.6 (\textcolor{mgreen}{+2.6}) \\
Qwen-2.5-Text (3B) &  68.3$\rightarrow$73.7 (\textcolor{mgreen}{+4.6})  &  76.6$\rightarrow$82.5 (\textcolor{mgreen}{+5.9}) \\
Qwen-2.5-Code (3B) &  71.2$\rightarrow$71.9 (\textcolor{mgreen}{+0.7})  &  80.3$\rightarrow$81.0 (\textcolor{mgreen}{+0.7}) \\
\bottomrule
\end{tabular}
}
\vspace{-0.9em}
\caption{\small Localization accuracy (Before$\rightarrow$After) from finetuning different listwise rerankers with \textsc{SweLoc}.}
\label{table:reranker_selection}
\vspace{-1em}
\end{wraptable}

Next, text- and code-instruction-tuned LLMs of different sizes from the Qwen2.5 family~\citep{Yang2024Qwen25TR, Hui2024Qwen25CoderTR} are finetuned as listwise LLM rerankers using \textsc{SweLoc} data. Since we only apply loss on the first generation token, to ensure compatibility with the listwise output format, all models were initially pre-trained on listwise text reranking data~\citep{pradeep2023rankzephyr}, which provides the full ranking order to use for supervision. The results, shown in Table~\ref{table:reranker_selection}, indicate that rerankers across different model sizes universally benefit from finetuning on \textsc{SweLoc}. An interesting observation is that the code-pretrained model performs marginally better at the 7B scale, while the text-pretrained models achieve better results at the 3B and 32B scales. Results with finetuning Llama-3.1 are in Appendix~\ref{sec:more_reranker}.

\subsubsection{Inference Cost Analysis}
\label{sec:cost_analysis}

\begin{wraptable}{r}{0.4\textwidth}
\vspace{-1em}
    \centering
    \small
    \renewcommand{\arraystretch}{1.35}
    \setlength{\tabcolsep}{3pt}
    \resizebox{\linewidth}{!}{
    \begin{tabular}{l l cc}
        \toprule
        \textbf{Method} & \textbf{Model} & \textbf{Cost(\$)} $\downarrow$ & $\displaystyle \frac{\textbf{Acc@10}}{\textbf{Cost}} \uparrow$\\
        \midrule
        \midrule
        \multirow{2}{*}{SWE-agent} & \texttt{GPT-4o} & 0.46 & 0.8 \\
         & \texttt{Claude-3.5} & 0.67 & 1.0 \\
        \midrule
        \multirow{2}{*}{Openhands} & \texttt{GPT-4o} & 0.83  & 0.6 \\
         & \texttt{Claude-3.5} & 0.79 & 0.9 \\
        \midrule
        \multirow{3}{*}{LocAgent} 
         & \texttt{Claude-3.5} & 0.66  & 1.2 \\
         &\texttt{Qwen2.5-7B(ft)} & 0.05  & 13.2 \\
         &\texttt{Qwen2.5-32B(ft)} & 0.09 & 8.6 \\
         \midrule
         \multirow{3}{*}{Reranker} &\texttt{GPT-4.1} & 0.16 & 5.9 \\
         & \textsc{SweRankLLM (7B)} & 0.011 &  \textbf{79.0} \\
         & \textsc{SweRankLLM (32B)} & 0.015 &  57.5 \\
        \bottomrule
    \end{tabular}
}
\vspace{-0.5em}
    \caption{\small \textsc{SweRankLLM} has considerably better inference cost-efficiency than agent-based methods while being more performant.}
    \label{tab:cost_efficiency}
    \vspace{-2em}
\end{wraptable}

Agent-based localization approaches typically involve multiple iterations, each requiring extensive chain-of-thought generation~\citep{wang2023selfconsistencyimproveschainthought}, incurring considerable cost at inference. In contrast, \textsc{SweRank} offers significant cost-effectiveness as the \textsc{SweRankLLM} reranker only needs to generate output candidate identifiers to determine the ranking order. Furthermore, the \textsc{SweRankEmbed} output embeddings can be pre-computed, resulting in negligible extra cost. Table~\ref{tab:cost_efficiency} compares the inference costs of \textsc{SweRankLLM} with other agent-based methods. Clearly, agent-based approaches, often relying on closed-source models for better performance, are highly cost-intensive. \textsc{SweRank} is substantially cheaper while providing significantly better performance, with up to \textit{6X} better performance-cost tradeoffs compared to LocAgent.

\subsubsection{Impact on Downstream Issue Resolution}
\label{sec:code_repair}

\begin{wraptable}{r}{0.35\linewidth}
\centering
\small
\vspace{-1.1em}
\renewcommand{\arraystretch}{1.35}
\setlength{\tabcolsep}{3pt}
\resizebox{\linewidth}{!}{%
\begin{tabular}{@{}l c c@{}}
\toprule
\textbf{Localization} & \textbf{File Acc@1} & \textbf{Repair Pass@1}\\ 
\midrule
SWE-Fixer & 69.7 & 21.0\\
LocAgent & 78.5 & 22.6 \\
SWERank & \textbf{83.2} & \textbf{24.5} \\ \hline
\demph{Oracle} & \demph{100} & \demph{25.9}\\
  \bottomrule
\end{tabular}
}
\vspace{-0.8em}
\caption{\small Impact of localization accuracy on downstream issue resolution.}
\label{tab:downstream_edit}    
\vspace{-1.5em}
\end{wraptable}
This section analyzes the impact of improved localization on downstream code repair performance. To evaluate issue resolution, we utilize SWE-Fixer~\citep{xie2025swe}, a two-step pipeline consisting of code file retrieval (localization) followed by code editing. We compare the repair outcomes on SWE-Bench-Lite when employing different localization methods: the native localization mechanism of SWE-Fixer, LocAgent (with Claude-3.5), our \textsc{SweRank} (large variant), and an oracle. The oracle simulates perfect localization by using the ground-truth edited file, thereby providing an upper bound for the repair framework. 
From Table~\ref{tab:downstream_edit}, we see that better localization provided by \textsc{SweRank} yields improved issue resolution, with oracle results showing that repair performance is currently constrained by the code editing model. 

\subsubsection{Performance Analysis by Issue Complexity}
\label{sec:complexity_analysis}

\begin{wrapfigure}{r}{0.6\linewidth}
\vspace{-1.5em}
    \centering
    \includegraphics[width=\linewidth]{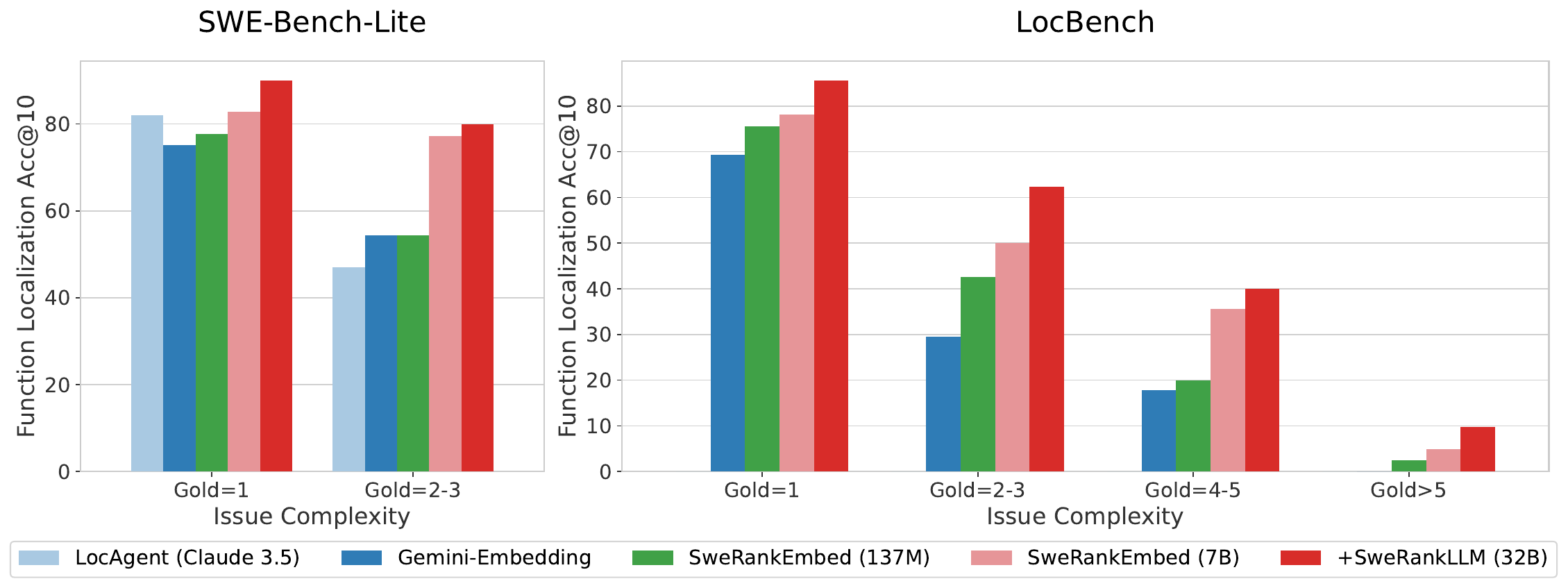}
    \vspace{-2em}
    \caption{\small Function Acc@10 breakdown by issue complexity.}
    \vspace{-1.7em}
    \label{fig:complexity_breakdown}
\end{wrapfigure}

Aggregate metrics often obscure performance variance on complex issues. To address this, we stratify the test set by \texttt{num\_gold} (the number of functions modified in the ground truth patch) as a proxy for issue complexity. 
We compare our approach against LocAgent and Gemini-Embedding, with results summarized in Figure~\ref{fig:complexity_breakdown}.


As expected, performance degrades as the number of modified functions increases. Instances requiring changes to $>5$ functions are extremely difficult for all models, resulting in single-digit accuracy. However, \textsc{SweRankEmbed-Large} demonstrates significantly better scaling on complex issues (\texttt{num\_gold}=2--3) compared to Agentic approaches; on SWE-Bench-Lite, LocAgent drops from 82.0\% to 47.1\%, while our retriever maintains 77.1\%. Furthermore, the reranker consistently improves performance across all complexity levels, confirming that it successfully captures cross-function dependencies that the bi-encoder might miss.

\subsubsection{Performance Analysis by Lexical and Semantic Overlap}
\label{sec:overlap_analysis}


To further dissect the model's capabilities, we analyzed performance by grouping instances based on \textbf{Lexical Overlap} (Rouge-1) and \textbf{Semantic Overlap} (Cosine Similarity).

\paragraph{Lexical Overlap.}
\begin{wrapfigure}{r}{0.65\linewidth}
\vspace{-1.25em}
    \centering
    \includegraphics[width=\linewidth]{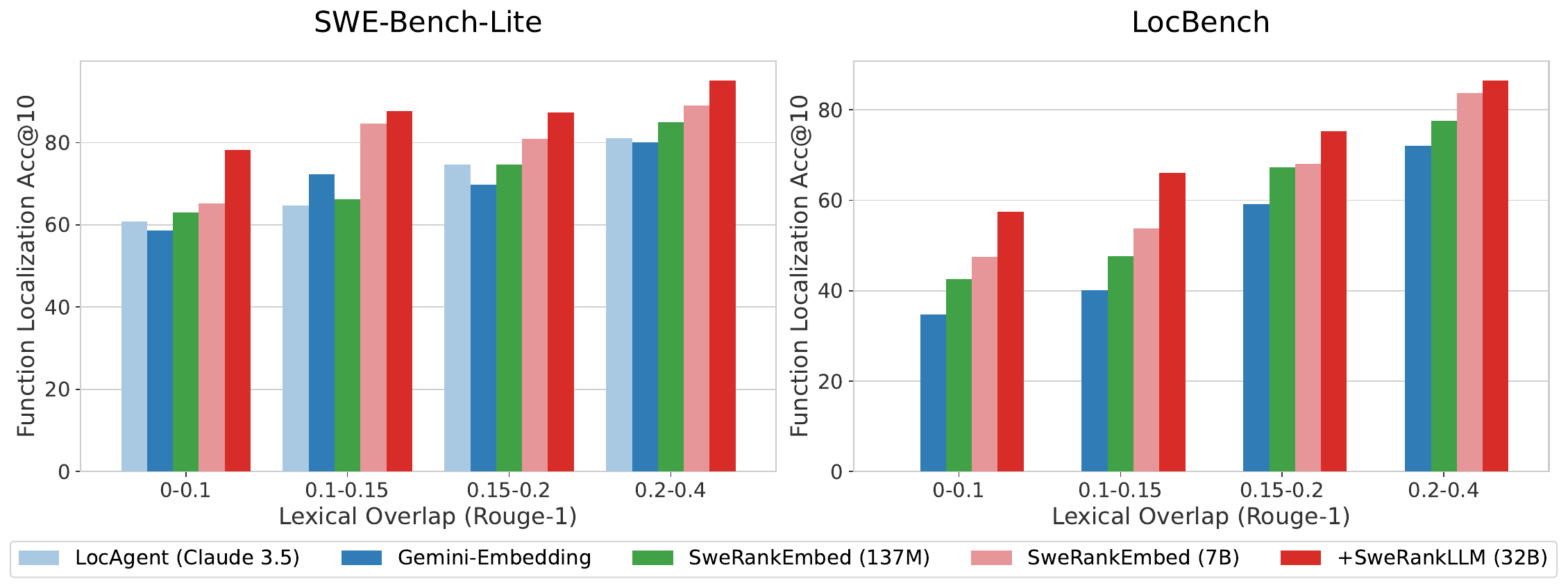}
    \vspace{-1.7em}
    \caption{\small Performance breakdown by \textbf{Lexical Overlap} (Rouge-1).}
    \vspace{-1.25em}
    \label{fig:lexical_overlap}
\end{wrapfigure}
We bucket instances into four groups using the Rouge-1 score between the issue description and the ground-truth localized functions. A high Rouge score indicates significant keyword overlap. Figure~\ref{fig:lexical_overlap} summarizes the results. We observe that performance generally degrades as lexical overlap decreases. However, even in the lowest overlap bucket (0.0--0.1), \textsc{SweRankEmbed-Large} outperforms LocAgent (65.2\% vs 60.9\% on SWE-Bench-Lite), demonstrating that our model does not rely solely on keyword matching. Furthermore, \textsc{SweRankLLM} consistently improves performance, with significant gains seen specifically for instances with low lexical overlap.

\paragraph{Semantic Overlap.}
\begin{wrapfigure}{r}{0.65\linewidth}
\vspace{-1.25em}
    \centering
    \includegraphics[width=\linewidth]{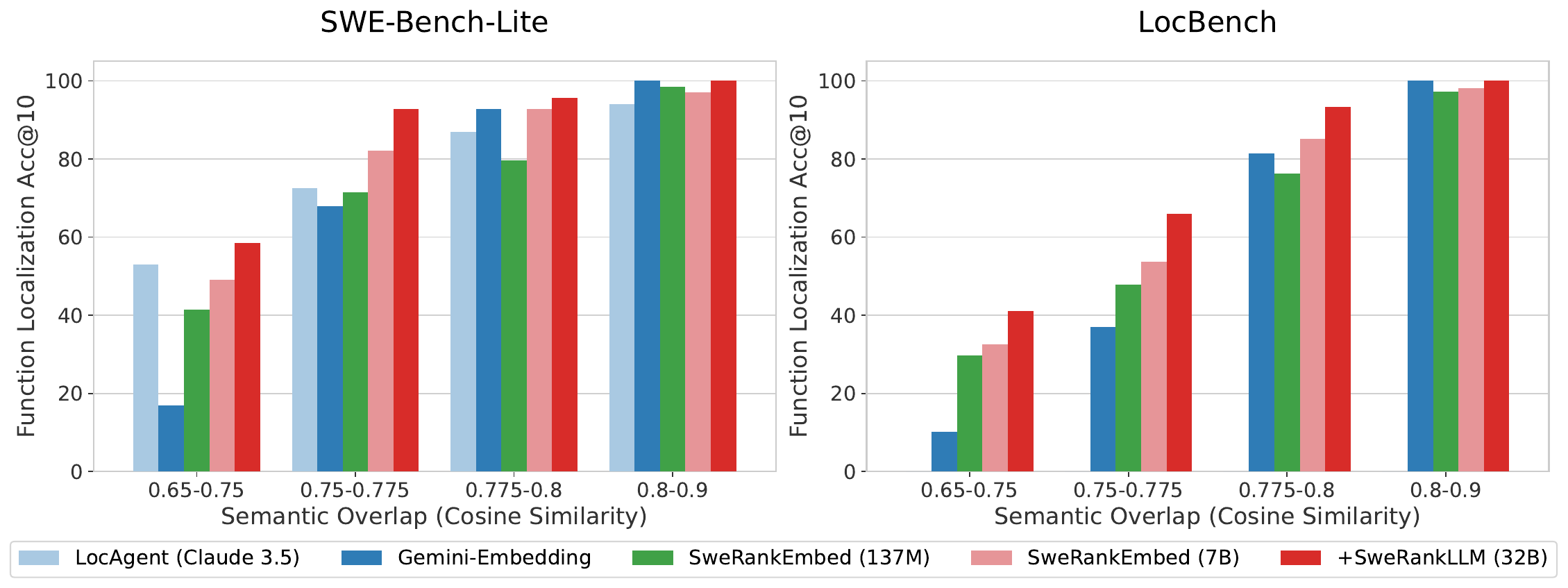}
    \vspace{-1.7em}
    \caption{\small Performance breakdown by \textbf{Semantic Overlap}.}
    \vspace{-1.25em}
    \label{fig:semantic_overlap}
\end{wrapfigure}
We also categorize instances based on the mean cosine similarity (computed via Gemini-Embedding) of the issue description and the ground-truth functions. As shown in Figure~\ref{fig:semantic_overlap}, performance is directly correlated with semantic overlap, achieving near-perfect accuracy for high-similarity instances ($>0.8$). Notably, \textsc{SweRankLLM} reranker considerably boosts performance over the retriever in low-similarity buckets (0.65--0.75), outperforming the multi-turn LocAgent approach. This suggests that while agentic tool-use can bridge the semantic gap, our training on \textsc{SweLoc}---which incorporates hard negatives---enables the \textsc{SweRank} framework to learn these non-obvious mappings effectively without the cost of agentic inference.

\section{Conclusion}

This paper frames software issue localization as a specialized ranking task and introduces \textsc{SweRank}, a highly performant and cost-effective retrieve-and-rerank framework. To effectively train \textsc{SweRank} models, we construct \textsc{SweLoc}, a large-scale contrastive training dataset derived from real-world GitHub issues, employing consistency filtering and hard-negative mining for quality. Empirical evaluations on SWE-Bench-Lite and LocBench demonstrate state-of-the-art localization performance using \textsc{SweRank}, significantly outperforming costly closed-source agent-based systems. 
The introduction of \textsc{SweLoc} dataset provides a valuable resource for advancing research in this domain.



\bibliography{iclr2026_conference}

\begin{thebibliography}{65}
\providecommand{\natexlab}[1]{#1}
\providecommand{\url}[1]{\texttt{#1}}
\expandafter\ifx\csname urlstyle\endcsname\relax
  \providecommand{\doi}[1]{doi: #1}\else
  \providecommand{\doi}{doi: \begingroup \urlstyle{rm}\Url}\fi

\bibitem[Adhiselvam et~al.(2015)Adhiselvam, Kirubakaran, and Sukumar]{adhiselvam2015enhanced}
A~Adhiselvam, E~Kirubakaran, and R~Sukumar.
\newblock An enhanced approach for software bug localization using map reduce technique based apriori (mrtba) algorithm.
\newblock \emph{Indian Journal of Science and Technology}, 8:\penalty0 35, 2015.

\bibitem[Amario~de Souza et~al.(2024)Amario~de Souza, de~Souza~Lauretto, Kon, and Lordello~Chaim]{amario2024understanding}
Higor Amario~de Souza, Marcelo de~Souza~Lauretto, Fabio Kon, and Marcos Lordello~Chaim.
\newblock Understanding the use of spectrum-based fault localization.
\newblock \emph{Journal of Software: Evolution and Process}, 36\penalty0 (6):\penalty0 e2622, 2024.

\bibitem[Anthropic(2023)]{anthropic2023claude}
Anthropic.
\newblock Claude: Conversational ai by anthropic, 2023.
\newblock URL \url{https://www.anthropic.com/claude}.
\newblock Accessed: January 21, 2025.

\bibitem[Billah et~al.(2024)Billah, Roy, Codabux, and Roy]{billah2024large}
Md~Mustakim Billah, Palash~Ranjan Roy, Zadia Codabux, and Banani Roy.
\newblock Are large language models a threat to programming platforms? an exploratory study.
\newblock In \emph{Proceedings of the 18th ACM/IEEE International Symposium on Empirical Software Engineering and Measurement}, pp.\  292--301, 2024.

\bibitem[Chen et~al.(2025)Chen, Tang, Deng, Wu, Wu, Jiang, Prasanna, Cohan, and Wang]{chen2025locagent}
Zhaoling Chen, Xiangru Tang, Gangda Deng, Fang Wu, Jialong Wu, Zhiwei Jiang, Viktor Prasanna, Arman Cohan, and Xingyao Wang.
\newblock Locagent: Graph-guided llm agents for code localization.
\newblock \emph{arXiv preprint arXiv:2503.09089}, 2025.

\bibitem[{Cognition AI}(2024)]{cognition2024devin}
{Cognition AI}.
\newblock {Devin: The First AI Software Engineer}.
\newblock \url{https://devin.ai/}, 2024.
\newblock Accessed: 2025-04-22.

\bibitem[{Cursor}(2025)]{cursor2025}
{Cursor}.
\newblock {Cursor: The AI Code Editor}.
\newblock \url{https://www.cursor.com/}, 2025.
\newblock Accessed: 2025-04-22.

\bibitem[de~Souza et~al.(2016)de~Souza, Chaim, and Kon]{de2016spectrum}
Higor~A de~Souza, Marcos~L Chaim, and Fabio Kon.
\newblock Spectrum-based software fault localization: A survey of techniques, advances, and challenges.
\newblock \emph{arXiv preprint arXiv:1607.04347}, 2016.

\bibitem[Elsaka(2017)]{ELSAKA201779}
E.~Elsaka.
\newblock Chapter three - fault localization using hybrid static/dynamic analysis.
\newblock volume 105 of \emph{Advances in Computers}, pp.\  79--114. Elsevier, 2017.
\newblock \doi{https://doi.org/10.1016/bs.adcom.2016.12.004}.
\newblock URL \url{https://www.sciencedirect.com/science/article/pii/S0065245816300778}.

\bibitem[Gauthier(2024)]{aider2024swe}
Paul Gauthier.
\newblock How aider scored sota 26.3\% on swe bench lite | aider, 2024.
\newblock URL \url{https://aider.chat/2024/05/22/swe-bench-lite.html}.
\newblock Accessed: January 21, 2025.

\bibitem[Grattafiori et~al.(2024)Grattafiori, Dubey, Jauhri, Pandey, Kadian, Al-Dahle, Letman, Mathur, Schelten, Vaughan, et~al.]{grattafiori2024llama}
Aaron Grattafiori, Abhimanyu Dubey, Abhinav Jauhri, Abhinav Pandey, Abhishek Kadian, Ahmad Al-Dahle, Aiesha Letman, Akhil Mathur, Alan Schelten, Alex Vaughan, et~al.
\newblock The llama 3 herd of models.
\newblock \emph{arXiv preprint arXiv:2407.21783}, 2024.

\bibitem[G{\"u}nther et~al.(2023)G{\"u}nther, Mastrapas, Wang, Xiao, and Geuter]{gunther2023jina}
Michael G{\"u}nther, Georgios Mastrapas, Bo~Wang, Han Xiao, and Jonathan Geuter.
\newblock Jina embeddings: A novel set of high-performance sentence embedding models.
\newblock In \emph{Proceedings of the 3rd Workshop for Natural Language Processing Open Source Software (NLP-OSS 2023)}, pp.\  8--18, 2023.

\bibitem[Günther et~al.(2023)Günther, Milliken, Geuter, Mastrapas, Wang, and Xiao]{günther2023jinaembeddingsnovelset}
Michael Günther, Louis Milliken, Jonathan Geuter, Georgios Mastrapas, Bo~Wang, and Han Xiao.
\newblock Jina embeddings: A novel set of high-performance sentence embedding models, 2023.
\newblock URL \url{https://arxiv.org/abs/2307.11224}.

\bibitem[Hui et~al.(2024)Hui, Yang, Cui, Yang, Liu, Zhang, Liu, Zhang, Yu, Dang, Yang, Men, Huang, Quan, Ren, Ren, Zhou, and Lin]{Hui2024Qwen25CoderTR}
Binyuan Hui, Jian Yang, Zeyu Cui, Jiaxi Yang, Dayiheng Liu, Lei Zhang, Tianyu Liu, Jiajun Zhang, Bowen Yu, Kai Dang, An~Yang, Rui Men, Fei Huang, Shanghaoran Quan, Xingzhang Ren, Xuancheng Ren, Jingren Zhou, and Junyang Lin.
\newblock Qwen2.5-coder technical report.
\newblock \emph{ArXiv}, abs/2409.12186, 2024.
\newblock URL \url{https://api.semanticscholar.org/CorpusID:272707390}.

\bibitem[Husain et~al.(2019)Husain, Wu, Gazit, Allamanis, and Brockschmidt]{husain2019codesearchnet}
Hamel Husain, Ho-Hsiang Wu, Tiferet Gazit, Miltiadis Allamanis, and Marc Brockschmidt.
\newblock Codesearchnet challenge: Evaluating the state of semantic code search.
\newblock \emph{arXiv preprint arXiv:1909.09436}, 2019.

\bibitem[Jain et~al.(2023)Jain, yeh Chiang, Wen, Kirchenbauer, Chu, Somepalli, Bartoldson, Kailkhura, Schwarzschild, Saha, Goldblum, Geiping, and Goldstein]{jain2023neftunenoisyembeddingsimprove}
Neel Jain, Ping yeh Chiang, Yuxin Wen, John Kirchenbauer, Hong-Min Chu, Gowthami Somepalli, Brian~R. Bartoldson, Bhavya Kailkhura, Avi Schwarzschild, Aniruddha Saha, Micah Goldblum, Jonas Geiping, and Tom Goldstein.
\newblock Neftune: Noisy embeddings improve instruction finetuning, 2023.
\newblock URL \url{https://arxiv.org/abs/2310.05914}.

\bibitem[Jimenez et~al.(2023)Jimenez, Yang, Wettig, Yao, Pei, Press, and Narasimhan]{jimenez2023swe}
Carlos~E Jimenez, John Yang, Alexander Wettig, Shunyu Yao, Kexin Pei, Ofir Press, and Karthik Narasimhan.
\newblock Swe-bench: Can language models resolve real-world github issues?
\newblock \emph{arXiv preprint arXiv:2310.06770}, 2023.

\bibitem[Jin et~al.(2024)Jin, Lee, Shin, and Kim]{jin2024teach}
Hyoungwook Jin, Seonghee Lee, Hyungyu Shin, and Juho Kim.
\newblock Teach ai how to code: Using large language models as teachable agents for programming education.
\newblock In \emph{Proceedings of the 2024 CHI Conference on Human Factors in Computing Systems}, pp.\  1--28, 2024.

\bibitem[Kocetkov et~al.(2022)Kocetkov, Li, Jia, Mou, Jernite, Mitchell, Ferrandis, Hughes, Wolf, Bahdanau, et~al.]{kocetkovstack}
Denis Kocetkov, Raymond Li, LI~Jia, Chenghao Mou, Yacine Jernite, Margaret Mitchell, Carlos~Mu{\~n}oz Ferrandis, Sean Hughes, Thomas Wolf, Dzmitry Bahdanau, et~al.
\newblock The stack: 3 tb of permissively licensed source code.
\newblock \emph{Transactions on Machine Learning Research}, 2022.

\bibitem[Kusupati et~al.(2022)Kusupati, Bhatt, Rege, Wallingford, Sinha, Ramanujan, Howard-Snyder, Chen, Kakade, Jain, and Farhadi]{matryoshak2022}
Aditya Kusupati, Gantavya Bhatt, Aniket Rege, Matthew Wallingford, Aditya Sinha, Vivek Ramanujan, William Howard-Snyder, Kaifeng Chen, Sham Kakade, Prateek Jain, and Ali Farhadi.
\newblock Matryoshka representation learning.
\newblock In S.~Koyejo, S.~Mohamed, A.~Agarwal, D.~Belgrave, K.~Cho, and A.~Oh (eds.), \emph{Advances in Neural Information Processing Systems}, volume~35, pp.\  30233--30249. Curran Associates, Inc., 2022.
\newblock URL \url{https://proceedings.neurips.cc/paper_files/paper/2022/file/c32319f4868da7613d78af9993100e42-Paper-Conference.pdf}.

\bibitem[Kwiatkowski et~al.(2019)Kwiatkowski, Palomaki, Redfield, Collins, Parikh, Alberti, Epstein, Polosukhin, Devlin, Lee, et~al.]{kwiatkowski2019natural}
Tom Kwiatkowski, Jennimaria Palomaki, Olivia Redfield, Michael Collins, Ankur Parikh, Chris Alberti, Danielle Epstein, Illia Polosukhin, Jacob Devlin, Kenton Lee, et~al.
\newblock Natural questions: a benchmark for question answering research.
\newblock \emph{Transactions of the Association for Computational Linguistics}, 7:\penalty0 453--466, 2019.

\bibitem[Lee et~al.(2025)Lee, Chen, Dua, Cer, Shanbhogue, Naim, {\'A}brego, Li, Chen, Vera, et~al.]{lee2025gemini}
Jinhyuk Lee, Feiyang Chen, Sahil Dua, Daniel Cer, Madhuri Shanbhogue, Iftekhar Naim, Gustavo~Hern{\'a}ndez {\'A}brego, Zhe Li, Kaifeng Chen, Henrique~Schechter Vera, et~al.
\newblock Gemini embedding: Generalizable embeddings from gemini.
\newblock \emph{arXiv preprint arXiv:2503.07891}, 2025.

\bibitem[Li et~al.(2024{\natexlab{a}})Li, Dong, Lee, Xia, Yin, Zhang, Liu, Wang, and Tang]{li2024codesearchnet}
Xiangyang Li, Kuicai Dong, Yi~Quan Lee, Wei Xia, Yichun Yin, Hao Zhang, Yong Liu, Yasheng Wang, and Ruiming Tang.
\newblock Csn: A comprehensive benchmark for code information retrieval models.
\newblock \emph{arXiv preprint arXiv:2407.02883}, 2024{\natexlab{a}}.

\bibitem[Li et~al.(2024{\natexlab{b}})Li, Dong, Lee, Xia, Yin, Zhang, Liu, Wang, and Tang]{li2024coir}
Xiangyang Li, Kuicai Dong, Yi~Quan Lee, Wei Xia, Yichun Yin, Hao Zhang, Yong Liu, Yasheng Wang, and Ruiming Tang.
\newblock Coir: A comprehensive benchmark for code information retrieval models.
\newblock \emph{arXiv preprint arXiv:2407.02883}, 2024{\natexlab{b}}.

\bibitem[Li et~al.(2023)Li, Zhang, Zhang, Long, Xie, and Zhang]{li2023towards}
Zehan Li, Xin Zhang, Yanzhao Zhang, Dingkun Long, Pengjun Xie, and Meishan Zhang.
\newblock Towards general text embeddings with multi-stage contrastive learning.
\newblock \emph{arXiv preprint arXiv:2308.03281}, 2023.

\bibitem[Liu et~al.(2024{\natexlab{a}})Liu, Xu, and McAuley]{liurepobench}
Tianyang Liu, Canwen Xu, and Julian McAuley.
\newblock Repobench: Benchmarking repository-level code auto-completion systems.
\newblock In \emph{The Twelfth International Conference on Learning Representations}, 2024{\natexlab{a}}.

\bibitem[Liu et~al.(2024{\natexlab{b}})Liu, Meng, Joty, Savarese, Xiong, Zhou, and Yavuz]{liu2024codexembed}
Ye~Liu, Rui Meng, Shafiq Joty, Silvio Savarese, Caiming Xiong, Yingbo Zhou, and Semih Yavuz.
\newblock Codexembed: A generalist embedding model family for multiligual and multi-task code retrieval.
\newblock \emph{arXiv preprint arXiv:2411.12644}, 2024{\natexlab{b}}.

\bibitem[Meng et~al.(2024)Meng, Liu, Jotya, Xiong, Zhou, and Yavuz]{SFR-embedding-2}
Rui Meng, Ye~Liu, Shafiq~Rayhan Jotya, Caiming Xiong, Yingbo Zhou, and Semih Yavuz.
\newblock Sfr-embedding-2: Advanced text embedding with multi-stage training, 2024.
\newblock URL \url{https://huggingface.co/Salesforce/SFR-Embedding-2_R}.

\bibitem[Merrick et~al.(2024)Merrick, Xu, Nuti, and Campos]{merrick2024arctic}
Luke Merrick, Danmei Xu, Gaurav Nuti, and Daniel Campos.
\newblock Arctic-embed: Scalable, efficient, and accurate text embedding models.
\newblock \emph{arXiv preprint arXiv:2405.05374}, 2024.

\bibitem[{Microsoft}(2023)]{microsoft2023copilot}
{Microsoft}.
\newblock {GitHub Copilot—Your AI pair programmer}, 2023.
\newblock URL \url{https://github.com/features/copilot}.

\bibitem[Moreira et~al.(2024)Moreira, Osmulski, Xu, Ak, Schifferer, and Oldridge]{moreira2024nv}
Gabriel de Souza~P Moreira, Radek Osmulski, Mengyao Xu, Ronay Ak, Benedikt Schifferer, and Even Oldridge.
\newblock Nv-retriever: Improving text embedding models with effective hard-negative mining.
\newblock \emph{arXiv preprint arXiv:2407.15831}, 2024.

\bibitem[Muennighoff et~al.(2023)Muennighoff, Tazi, Magne, and Reimers]{muennighoff2023mteb}
Niklas Muennighoff, Nouamane Tazi, Loic Magne, and Nils Reimers.
\newblock Mteb: Massive text embedding benchmark.
\newblock In \emph{Proceedings of the 17th Conference of the European Chapter of the Association for Computational Linguistics}, pp.\  2014--2037, 2023.

\bibitem[Nussbaum et~al.(2024)Nussbaum, Morris, Duderstadt, and Mulyar]{nussbaum2024nomic}
Zach Nussbaum, John~X. Morris, Brandon Duderstadt, and Andriy Mulyar.
\newblock Nomic embed: Training a reproducible long context text embedder, 2024.

\bibitem[Oord et~al.(2018)Oord, Li, and Vinyals]{oord2018representation}
Aaron van~den Oord, Yazhe Li, and Oriol Vinyals.
\newblock Representation learning with contrastive predictive coding.
\newblock In \emph{Advances in Neural Information Processing Systems}, pp.\  10203--10213, 2018.

\bibitem[Pan et~al.(2024)Pan, Ibrahimzada, Krishna, Sankar, Wassi, Merler, Sobolev, Pavuluri, Sinha, and Jabbarvand]{pan2024lost}
Rangeet Pan, Ali~Reza Ibrahimzada, Rahul Krishna, Divya Sankar, Lambert~Pouguem Wassi, Michele Merler, Boris Sobolev, Raju Pavuluri, Saurabh Sinha, and Reyhaneh Jabbarvand.
\newblock Lost in translation: A study of bugs introduced by large language models while translating code.
\newblock In \emph{Proceedings of the IEEE/ACM 46th International Conference on Software Engineering}, pp.\  1--13, 2024.

\bibitem[Pradeep et~al.(2023{\natexlab{a}})Pradeep, Sharifymoghaddam, and Lin]{pradeep2023rankvicunazeroshotlistwisedocument}
Ronak Pradeep, Sahel Sharifymoghaddam, and Jimmy Lin.
\newblock Rankvicuna: Zero-shot listwise document reranking with open-source large language models, 2023{\natexlab{a}}.
\newblock URL \url{https://arxiv.org/abs/2309.15088}.

\bibitem[Pradeep et~al.(2023{\natexlab{b}})Pradeep, Sharifymoghaddam, and Lin]{pradeep2023rankzephyr}
Ronak Pradeep, Sahel Sharifymoghaddam, and Jimmy Lin.
\newblock Rankzephyr: Effective and robust zero-shot listwise reranking is a breeze!
\newblock \emph{arXiv preprint arXiv:2312.02724}, 2023{\natexlab{b}}.

\bibitem[Qin et~al.(2024)Qin, Wang, Lou, Dong, Wang, Li, and Mao]{qin2024agentfl}
Yihao Qin, Shangwen Wang, Yiling Lou, Jinhao Dong, Kaixin Wang, Xiaoling Li, and Xiaoguang Mao.
\newblock Agentfl: Scaling llm-based fault localization to project-level context.
\newblock \emph{arXiv preprint arXiv:2403.16362}, 2024.

\bibitem[Rasley et~al.(2020)Rasley, Rajbhandari, Ruwase, and He]{rasley2020deepspeed}
Jeff Rasley, Samyam Rajbhandari, Olatunji Ruwase, and Yuxiong He.
\newblock Deepspeed: System optimizations enable training deep learning models with over 100 billion parameters.
\newblock In \emph{Proceedings of the 26th ACM SIGKDD International Conference on Knowledge Discovery \& Data Mining}, pp.\  3505--3506, 2020.

\bibitem[Reddy et~al.(2024)Reddy, Doo, Xu, Sultan, Swain, Sil, and Ji]{reddy2024first}
Revanth~Gangi Reddy, JaeHyeok Doo, Yifei Xu, Md~Arafat Sultan, Deevya Swain, Avirup Sil, and Heng Ji.
\newblock First: Faster improved listwise reranking with single token decoding.
\newblock In \emph{Proceedings of the 2024 Conference on Empirical Methods in Natural Language Processing}, pp.\  8642--8652, 2024.

\bibitem[Reimers \& Gurevych(2019)Reimers and Gurevych]{reimers-gurevych-2019-sentence}
Nils Reimers and Iryna Gurevych.
\newblock Sentence-{BERT}: Sentence embeddings using {S}iamese {BERT}-networks.
\newblock In Kentaro Inui, Jing Jiang, Vincent Ng, and Xiaojun Wan (eds.), \emph{Proceedings of the 2019 Conference on Empirical Methods in Natural Language Processing and the 9th International Joint Conference on Natural Language Processing (EMNLP-IJCNLP)}, pp.\  3982--3992, Hong Kong, China, November 2019. Association for Computational Linguistics.
\newblock \doi{10.18653/v1/D19-1410}.
\newblock URL \url{https://aclanthology.org/D19-1410}.

\bibitem[Robertson et~al.(1994)Robertson, Walker, Jones, Hancock-Beaulieu, and Gatford]{robertson1994okapi}
Stephen~E. Robertson, Steve Walker, Susan Jones, Micheline Hancock-Beaulieu, and Mike Gatford.
\newblock Okapi at trec-3.
\newblock In \emph{Text Retrieval Conference}, 1994.
\newblock URL \url{https://api.semanticscholar.org/CorpusID:41563977}.

\bibitem[Sontakke et~al.(2022)Sontakke, Patwardhan, Vig, Medicherla, Naik, and Shroff]{sontakke2022code}
Ankita~Nandkishor Sontakke, Manasi Patwardhan, Lovekesh Vig, Raveendra~Kumar Medicherla, Ravindra Naik, and Gautam Shroff.
\newblock Code summarization: Do transformers really understand code?
\newblock In \emph{Deep Learning for Code Workshop}, 2022.

\bibitem[Soremekun et~al.(2021)Soremekun, Kirschner, B{\"o}hme, and Zeller]{soremekun2021locating}
Ezekiel Soremekun, Lukas Kirschner, Marcel B{\"o}hme, and Andreas Zeller.
\newblock Locating faults with program slicing: an empirical analysis.
\newblock \emph{Empirical Software Engineering}, 26:\penalty0 1--45, 2021.

\bibitem[Suresh et~al.(2024)Suresh, Reddy, Xu, Nussbaum, Mulyar, Duderstadt, and Ji]{suresh2024cornstack}
Tarun Suresh, Revanth~Gangi Reddy, Yifei Xu, Zach Nussbaum, Andriy Mulyar, Brandon Duderstadt, and Heng Ji.
\newblock Cornstack: High-quality contrastive data for better code ranking.
\newblock \emph{arXiv preprint arXiv:2412.01007}, 2024.

\bibitem[Thorne et~al.(2018)Thorne, Vlachos, Christodoulopoulos, and Mittal]{Thorne18Fever}
James Thorne, Andreas Vlachos, Christos Christodoulopoulos, and Arpit Mittal.
\newblock {FEVER}: a large-scale dataset for fact extraction and {VERification}.
\newblock In \emph{NAACL-HLT}, 2018.

\bibitem[Wang et~al.(2023{\natexlab{a}})Wang, Wang, Joty, and Hoi]{Wang-FSE23}
Weishi Wang, Yue Wang, Shafiq Joty, and Steven~C.H. Hoi.
\newblock Rap-gen: Retrieval-augmented patch generation with codet5 for automatic program repair.
\newblock In \emph{Proceedings of the 31st ACM Joint European Software Engineering Conference and Symposium on the Foundations of Software Engineering}, ESEC/FSE 2023, pp.\  146–158, New York, NY, USA, 2023{\natexlab{a}}. Association for Computing Machinery.
\newblock ISBN 9798400703270.
\newblock \doi{10.1145/3611643.3616256}.
\newblock URL \url{https://doi.org/10.1145/3611643.3616256}.

\bibitem[Wang et~al.(2025)Wang, Li, Song, Xu, Tang, Zhuge, Pan, Song, Li, Singh, Tran, Li, Ma, Zheng, Qian, Shao, Muennighoff, Zhang, Hui, Lin, Brennan, Peng, Ji, and Neubig]{wang2024openhands}
Xingyao Wang, Boxuan Li, Yufan Song, Frank~F. Xu, Xiangru Tang, Mingchen Zhuge, Jiayi Pan, Yueqi Song, Bowen Li, Jaskirat Singh, Hoang~H. Tran, Fuqiang Li, Ren Ma, Mingzhang Zheng, Bill Qian, Yanjun Shao, Niklas Muennighoff, Yizhe Zhang, Binyuan Hui, Junyang Lin, Robert Brennan, Hao Peng, Heng Ji, and Graham Neubig.
\newblock Openhands: An open platform for {AI} software developers as generalist agents.
\newblock In \emph{The Thirteenth International Conference on Learning Representations}, 2025.
\newblock URL \url{https://openreview.net/forum?id=OJd3ayDDoF}.

\bibitem[Wang et~al.(2023{\natexlab{b}})Wang, Wei, Schuurmans, Le, Chi, Narang, Chowdhery, and Zhou]{wang2023selfconsistencyimproveschainthought}
Xuezhi Wang, Jason Wei, Dale Schuurmans, Quoc Le, Ed~Chi, Sharan Narang, Aakanksha Chowdhery, and Denny Zhou.
\newblock Self-consistency improves chain of thought reasoning in language models, 2023{\natexlab{b}}.
\newblock URL \url{https://arxiv.org/abs/2203.11171}.

\bibitem[Wang et~al.(2023{\natexlab{c}})Wang, Le, Gotmare, Bui, Li, and Hoi]{wang2023codet5opencodelarge}
Yue Wang, Hung Le, Akhilesh~Deepak Gotmare, Nghi D.~Q. Bui, Junnan Li, and Steven C.~H. Hoi.
\newblock Codet5+: Open code large language models for code understanding and generation, 2023{\natexlab{c}}.
\newblock URL \url{https://arxiv.org/abs/2305.07922}.

\bibitem[{Windsurf}(2025)]{windsurf2025}
{Windsurf}.
\newblock {Windsurf Editor: The AI‑Native IDE}.
\newblock \url{https://windsurf.com/editor}, 2025.
\newblock Accessed: 2025-04-22.

\bibitem[Wong et~al.(2016)Wong, Gao, Li, Abreu, and Wotawa]{wong2016survey}
W~Eric Wong, Ruizhi Gao, Yihao Li, Rui Abreu, and Franz Wotawa.
\newblock A survey on software fault localization.
\newblock \emph{IEEE Transactions on Software Engineering}, 42\penalty0 (8):\penalty0 707--740, 2016.

\bibitem[Xie et~al.(2025)Xie, Li, Gao, Du, Lam, Zou, and Chen]{xie2025swe}
Chengxing Xie, Bowen Li, Chang Gao, He~Du, Wai Lam, Difan Zou, and Kai Chen.
\newblock Swe-fixer: Training open-source llms for effective and efficient github issue resolution.
\newblock \emph{arXiv preprint arXiv:2501.05040}, 2025.

\bibitem[Yang et~al.(2024{\natexlab{a}})Yang, Jimenez, Wettig, Lieret, Yao, Narasimhan, and Press]{yang2024swe}
John Yang, Carlos Jimenez, Alexander Wettig, Kilian Lieret, Shunyu Yao, Karthik Narasimhan, and Ofir Press.
\newblock Swe-agent: Agent-computer interfaces enable automated software engineering.
\newblock \emph{Advances in Neural Information Processing Systems}, 37:\penalty0 50528--50652, 2024{\natexlab{a}}.

\bibitem[Yang et~al.(2024{\natexlab{b}})Yang, Jimenez, Wettig, Lieret, Yao, Narasimhan, and Press]{yang2024sweagent}
John Yang, Carlos~E Jimenez, Alexander Wettig, Kilian Lieret, Shunyu Yao, Karthik Narasimhan, and Ofir Press.
\newblock Swe-agent: Agent-computer interfaces enable automated software engineering.
\newblock \emph{arXiv preprint arXiv:2405.15793}, 2024{\natexlab{b}}.

\bibitem[Yang et~al.(2025)Yang, Lieret, Jimenez, Wettig, Khandpur, Zhang, Hui, Press, Schmidt, and Yang]{yang2025swesmith}
John Yang, Kilian Lieret, Carlos~E. Jimenez, Alexander Wettig, Kabir Khandpur, Yanzhe Zhang, Binyuan Hui, Ofir Press, Ludwig Schmidt, and Diyi Yang.
\newblock Swe-smith: Scaling data for software engineering agents, 2025.
\newblock URL \url{https://arxiv.org/abs/2504.21798}.

\bibitem[Yang et~al.(2024{\natexlab{c}})Yang, Yang, Zhang, Hui, Zheng, Yu, Li, Liu, Huang, Dong, Wei, Lin, Yang, Tu, Zhang, Yang, Yang, Zhou, Lin, Dang, Lu, Bao, Yang, Yu, Li, Xue, Zhang, Zhu, Men, Lin, Li, Xia, Ren, Ren, Fan, Su, Zhang, Wan, Liu, Cui, Zhang, Qiu, Quan, and Wang]{Yang2024Qwen25TR}
Qwen~An Yang, Baosong Yang, Beichen Zhang, Binyuan Hui, Bo~Zheng, Bowen Yu, Chengyuan Li, Dayiheng Liu, Fei Huang, Guanting Dong, Haoran Wei, Huan Lin, Jian Yang, Jianhong Tu, Jianwei Zhang, Jianxin Yang, Jiaxin Yang, Jingren Zhou, Junyang Lin, Kai Dang, Keming Lu, Keqin Bao, Kexin Yang, Le~Yu, Mei Li, Mingfeng Xue, Pei Zhang, Qin Zhu, Rui Men, Runji Lin, Tianhao Li, Tingyu Xia, Xingzhang Ren, Xuancheng Ren, Yang Fan, Yang Su, Yi-Chao Zhang, Yunyang Wan, Yuqi Liu, Zeyu Cui, Zhenru Zhang, Zihan Qiu, Shanghaoran Quan, and Zekun Wang.
\newblock Qwen2.5 technical report.
\newblock \emph{ArXiv}, abs/2412.15115, 2024{\natexlab{c}}.
\newblock URL \url{https://api.semanticscholar.org/CorpusID:274859421}.

\bibitem[Yang et~al.(2018)Yang, Qi, Zhang, Bengio, Cohen, Salakhutdinov, and Manning]{yang2018hotpotqa}
Zhilin Yang, Peng Qi, Saizheng Zhang, Yoshua Bengio, William Cohen, Ruslan Salakhutdinov, and Christopher~D Manning.
\newblock Hotpotqa: A dataset for diverse, explainable multi-hop question answering.
\newblock In \emph{Proceedings of the 2018 Conference on Empirical Methods in Natural Language Processing}. Association for Computational Linguistics, 2018.

\bibitem[Yao et~al.(2023)Yao, Zhao, Yu, Du, Shafran, Narasimhan, and Cao]{yao2023react}
Shunyu Yao, Jeffrey Zhao, Dian Yu, Nan Du, Izhak Shafran, Karthik Narasimhan, and Yuan Cao.
\newblock React: Synergizing reasoning and acting in language models.
\newblock In \emph{International Conference on Learning Representations (ICLR)}, 2023.

\bibitem[Yu et~al.(2024)Yu, Merrick, Nuti, and Campos]{yu2024arctic}
Puxuan Yu, Luke Merrick, Gaurav Nuti, and Daniel Campos.
\newblock Arctic-embed 2.0: Multilingual retrieval without compromise.
\newblock \emph{arXiv preprint arXiv:2412.04506}, 2024.

\bibitem[Yu et~al.(2025)Yu, Zhang, Zhao, Huang, Yao, Ding, and Zhao]{yu2025orcaloca}
Zhongming Yu, Hejia Zhang, Yujie Zhao, Hanxian Huang, Matrix Yao, Ke~Ding, and Jishen Zhao.
\newblock Orcaloca: An llm agent framework for software issue localization.
\newblock \emph{arXiv preprint arXiv:2502.00350}, 2025.

\bibitem[Yue et~al.(2021)Yue, Weishi, Joty, and Hoi]{wang2021codet5}
Wang Yue, Wang Weishi, Shafiq Joty, and Steven~C.H. Hoi.
\newblock Codet5: Identifier-aware unified pre-trained encoder-decoder models for code understanding and generation.
\newblock In \emph{EMNLP}, 2021.

\bibitem[Zhang et~al.(2024)Zhang, Ahmad, Tan, Ding, Nallapati, Roth, Ma, and Xiang]{zhang2024code}
Dejiao Zhang, Wasi~Uddin Ahmad, Ming Tan, Hantian Ding, Ramesh Nallapati, Dan Roth, Xiaofei Ma, and Bing Xiang.
\newblock {CODE} {REPRESENTATION} {LEARNING} {AT} {SCALE}.
\newblock In \emph{The Twelfth International Conference on Learning Representations}, 2024.
\newblock URL \url{https://openreview.net/forum?id=vfzRRjumpX}.

\bibitem[Zhang et~al.(2025)Zhang, Li, Long, Zhang, Lin, Yang, Xie, Yang, Liu, Lin, Huang, and Zhou]{zhang2025qwen3embedding}
Yanzhao Zhang, Mingxin Li, Dingkun Long, Xin Zhang, Huan Lin, Baosong Yang, Pengjun Xie, An~Yang, Dayiheng Liu, Junyang Lin, Fei Huang, and Jingren Zhou.
\newblock Qwen3 embedding: Advancing text embedding and reranking through foundation models, 2025.
\newblock URL \url{https://arxiv.org/abs/2506.05176}.

\bibitem[Örwall(2024)]{orwall2023moatless}
Albert Örwall.
\newblock Moatless tools, 2024.
\newblock URL \url{https://github.com/aorwall/moatless-tools}.

\end{thebibliography}
\bibliographystyle{iclr2026_conference}
\clearpage
\appendix



\section{Training Details}
\label{sec:model_training}
\subsection{\textsc{SweRankEmbed}}

 Our data filtering, negative mining, and model finetuning
are implemented using the contrastors package~\citep{nussbaum2024nomic}. The \textsc{SweRankEmbed-Small} encoder uses \textsc{CodeRankEmbed}, which was initialized with Arctic-Embed-M~\citep{merrick2024arctic}, a text encoder supporting an extended context length of 8,192 tokens and pretrained on large-scale web query-
document pairs, along with public text retrieval datasets~\citep{yang2018hotpotqa, kwiatkowski2019natural, Thorne18Fever}. The encoder supports a query prefix  ``\textit{Represent this query for searching relevant code: }'', as set by~\citep{suresh2024cornstack}. The model is finetuned using 8 GH200 GPUs for two epochs with a learning rate of 2e-5, a batch size of 64 and 15 hard negatives per example. 

The \textsc{SweRankEmbed-Large} encoder uses GTE-Qwen2-7B-Instruct~\citep{li2023towards}, which was pretrained on a large corpora of text retrieval data. For this model, we use a custom query prefix ``\textit{Instruct: Given a github issue, identify the code that needs to be changed to fix the issue. Query: }''. The model is finetuned using 8 GH200 GPUs for 1 epoch with a learning rate of 8e-6, a batch size
of 64 and 7 hard negatives per example. 

\subsection{\textsc{SweRankLLM}}
\paragraph{Training data:} For each $<$query, positive, negatives$>$ tuple from \textsc{SweLoc}, we randomly sample 9 negative codes to fit the listwise reranking window size of 10 along with the positive code. To prevent the positional bias from affecting the reranker and ensure model robustness~\citep{pradeep2023rankvicunazeroshotlistwisedocument}, we shuffle the order of candidate codes for each training example. Since the combined length of a GitHub issue and corresponding candidate codes may exceed the model's maximum embedding size, we set the maximum length per candidate code to 1024 and the total length limit to 16348. For overlong prompts, we truncate the query to reach the maximum total length. This preserves meaningful context for issue localization as much as possible within the limited context window size for effective model training. The rerankers are all first pretrained with text listwise reranking data~\citep{pradeep2023rankzephyr} to teach the model to follow the listwise output format.


\paragraph{Hyperparameters:} 
For the LLM reranker training, with both text reranking and \textsc{SweLoc} data, we trained for one epoch with a global batch size of 128, an initial learning rate of 5e-6 with 50 warmup steps, cosine learning rate scheduler, bfloat16 precision, and noisy embeddings~\citep{jain2023neftunenoisyembeddingsimprove} with a noise scale $\alpha = 5$. For efficient long-context, multi-gpu training, we used DeepSpeed~\citep{rasley2020deepspeed} ZeRO stage 3 with 16 GH200 GPUs.

\section{Experiments with More Reranker Models}
\label{sec:more_reranker}

\begin{wraptable}{r}{0.5\textwidth}
\vspace{-1em}
\centering
\small
\renewcommand{\arraystretch}{1.35}
\setlength{\tabcolsep}{3pt}
\resizebox{\linewidth}{!}{
\begin{tabular}{@{}lcccc@{}}
\toprule
\multirow{2}{*}{\textbf{Method Type}} & \multicolumn{2}{c}{\textbf{SWE-Bench-Lite}} & \multicolumn{2}{c}{\textbf{LocBench}}\\
 \cmidrule(lr){2-3}  \cmidrule(lr){4-5}
& \textbf{Acc@5}  & \textbf{Acc@10} & \textbf{Acc@5}  & \textbf{Acc@10} \\
\midrule
Zeroshot Reranker & 	60.22 & 81.39 & 61.96 & 69.11 \\
RankZephyr finetune & 72.99 & 80.29 & 64.11 & 70.00\\
+ \textsc{SweLoc} finetune & 	\textbf{77.01} & \textbf{85.77} & \textbf{68.04} & \textbf{73.04}\\
\bottomrule
\end{tabular}
}
\vspace{-0.6em}
\caption{Function localization accuracy of Llama-3.1 8B Instruct as a listwise LLM reranker.}
\label{table:more_reranker}
\vspace{-1em}
\end{wraptable}

To demonstrate the broader applicability of our dataset, we conduct experiments with finetuning Llama-3.1 8B Instruct~\citep{grattafiori2024llama} as a listwise reranker. The models are first pre-trained on general text reranking data from RankZephyr~\citep{pradeep2023rankzephyr} and subsequently finetuned on our \textsc{SweLoc} dataset. Results, shown in Table~\ref{table:more_reranker}, demonstrate significant performance gains on both SWE-Bench-Lite and LocBench after fine-tuning on SWELoc. This confirms that our dataset is a valuable resource for improving the issue localization capabilities of various LLM families, not just Qwen 2.5.


\section{Retriever Ceiling Analysis}

To assess the upper bound (performance ceiling) provided by the retrieval stage, we report extended metrics (Acc@20, @50, and @100) in Table~\ref{tab:extended_retrieval}. We compare our models against Gemini-Embedding~\citep{lee2025gemini}. On SWE-Bench-Lite, \textsc{SweRankEmbed-Large} achieves a retrieval ceiling (Acc@100) of 93.43\%. Given that \textsc{SweRankLLM-Large} achieves 88.7\% Acc@10, the gap suggests that the retrieval stage is not the primary bottleneck. Furthermore, this retrieval ceiling is significantly higher than the best performance achieved by agentic methods like LocAgent ($\sim$78\%).

\begin{table*}[t]
    \centering
    \small
    \renewcommand{\arraystretch}{1.25}
    \setlength{\tabcolsep}{6pt}
    \resizebox{\textwidth}{!}{%
    \begin{tabular}{@{}l cccc c cccc@{}}
        \toprule
        \multirow{2.5}{*}{\textbf{Model}} & \multicolumn{4}{c}{\textbf{SWE-Bench-Lite} (Acc@$K$)} & & \multicolumn{4}{c}{\textbf{LocBench} (Acc@$K$)} \\
        \cmidrule{2-5} \cmidrule{7-10}
        & \textbf{Acc@10} & \textbf{Acc@20} & \textbf{Acc@50} & \textbf{Acc@100} & & \textbf{Acc@10} & \textbf{Acc@20} & \textbf{Acc@50} & \textbf{Acc@100} \\
        \midrule
        SweRankEmbed-Small & 74.45 & 81.75 & 87.96 & 91.97 && 58.57 & 67.50 & 75.71 & 82.32 \\
        SweRankEmbed-Large & \textbf{82.12} & \textbf{86.50} & \textbf{90.88} & \textbf{93.43} && \textbf{63.21} & \textbf{71.25} & \textbf{80.71} & \textbf{84.29} \\
        \rowcolor{gray!10} Gemini-Embedding & 72.26 & 79.20 & 87.96 & 90.88 && 51.43 & 60.18 & 70.00 & 78.39 \\
        \bottomrule
    \end{tabular}%
    }
    \caption{Extended retrieval metrics. Acc@$K$ indicates the percentage of instances where all ground-truth functions are within the top-$K$ retrieved candidates. The Acc@100 score indicates a performance ceiling for the subsequent reranker.}
    \label{tab:extended_retrieval}
\end{table*}

\section{Ablation Studies on Model Capacity}
\label{sec:appendix_capacity}

While the experimental results in the main text demonstrates that performance generally improves with model size for rerankers, we provide additional experiments here to analyze the sensitivity of the retriever to model capacity and architecture design.

\paragraph{Impact of Encoder Depth \& Width:}
To isolate the effects of model architecture, we compare our \textsc{SweRankEmbed} variants against the recently released Qwen3-Embedding models~\citep{zhang2025qwen3embedding} (0.6B and 8B variants). We finetuned the Qwen3 models on the \textsc{SweLoc} dataset using the exact same procedure as \textsc{SweRankEmbed}. Table~\ref{tab:capacity_ablation} details the model specifications and performance. Comparing \textsc{SweRankEmbed-Small} to Qwen3-0.6B, we observe moderate gains from utilizing a significantly deeper and wider encoder. However, comparing \textsc{SweRankEmbed-Large} to Qwen3-8B suggests diminishing returns from further increasing depth (28 vs. 36 layers). Conversely, the considerable performance gap between Qwen3-0.6B and \textsc{SweRankEmbed-Large} appears driven by the larger embedding dimension, which we investigate below.

\begin{table*}[t]
    \centering
    \small
    \renewcommand{\arraystretch}{1.25}
    \setlength{\tabcolsep}{10pt} 
    \resizebox{\textwidth}{!}{%
    \begin{tabular}{@{}l ccc cc@{}}
        \toprule
        \textbf{Model} & \textbf{Depth} & \textbf{Width} & \textbf{Dim} & \textbf{Acc@5} & \textbf{Acc@10} \\
        \midrule
        SweRankEmbed-Small (137M) & 12 & 768 & 768 & 51.82 $\to$ 63.14 & 58.76 $\to$ 74.45 \\
        Qwen3-Embedding-0.6B & 28 & 2048 & 1024 & 52.55 $\to$ 66.79 & 62.77 $\to$ 75.18 \\
        SweRankEmbed-Large (7B) & 28 & 3072 & 3584 & 63.14 $\to$ 71.90 & 70.44 $\to$ 82.12 \\
        Qwen3-Embedding-8B & 36 & 3072 & 4096 & 60.95 $\to$ \textbf{73.72} & 71.53 $\to$ \textbf{83.94} \\
        \bottomrule
    \end{tabular}%
    }
    \caption{Impact of retriever model capacity. We compare different variants of the \textsc{SweRankEmbed} and Qwen3-Embedding models. Performance is reported as (Before $\to$ After) finetuning on \textsc{SweLoc}.}
    \vspace{-1em}
    \label{tab:capacity_ablation}
\end{table*}

\paragraph{Impact of Embedding Dimension:}
To strictly isolate the impact of embedding dimension, we performed a controlled ablation using the Qwen3-Embedding-0.6B model, which supports flexible vector dimensions via Matryoshka Representation Learning (MRL)~\citep{matryoshak2022}. Table~\ref{tab:dimension_ablation} presents the results on SWE-Bench-Lite after finetuning with MRL.

\begin{table}[h]
    \centering
    \small
    \renewcommand{\arraystretch}{1.25}
    \setlength{\tabcolsep}{10pt}
    \begin{tabular}{@{}c cc@{}}
        \toprule
        \textbf{Dimension ($D$)} & \textbf{Acc@5} (Before $\to$ After) & \textbf{Acc@10} (Before $\to$ After) \\
        \midrule
        1024 & 52.55 $\to$ \textbf{66.79} & 62.77 $\to$ \textbf{75.18} \\
        512 & 50.00 $\to$ 65.69 & 59.85 $\to$ 74.09 \\
        256 & 44.16 $\to$ 59.12 & 52.92 $\to$ 69.71 \\
        128 & 39.05 $\to$ 56.93 & 46.72 $\to$ 66.79 \\
        64 & 33.21 $\to$ 50.00 & 38.32 $\to$ 60.22 \\
        \bottomrule
    \end{tabular}
    \caption{Controlled ablation on embedding dimension using Qwen3-Embedding-0.6B with Matryoshka Representation Learning. \textsc{SweLoc} provides larger relative gains at lower dimensions.}
    \label{tab:dimension_ablation}
\end{table}

Performance drops significantly as embedding size decreases, identifying dimension as a critical factor. Interestingly, finetuning on \textsc{SweLoc} yields larger relative gains at lower dimensions (e.g., +21.9 points Acc@10 for $D=64$ vs. +12.5 points for $D=1024$), highlighting the dataset's utility even for compressed representations.

\section{Benefit of Iterative Negative Mining}
\label{sec:iterative_mining}

\begin{wrapfigure}{r}{0.48\linewidth}
\vspace{-1.25em}
\centering
\includegraphics[width=\linewidth]{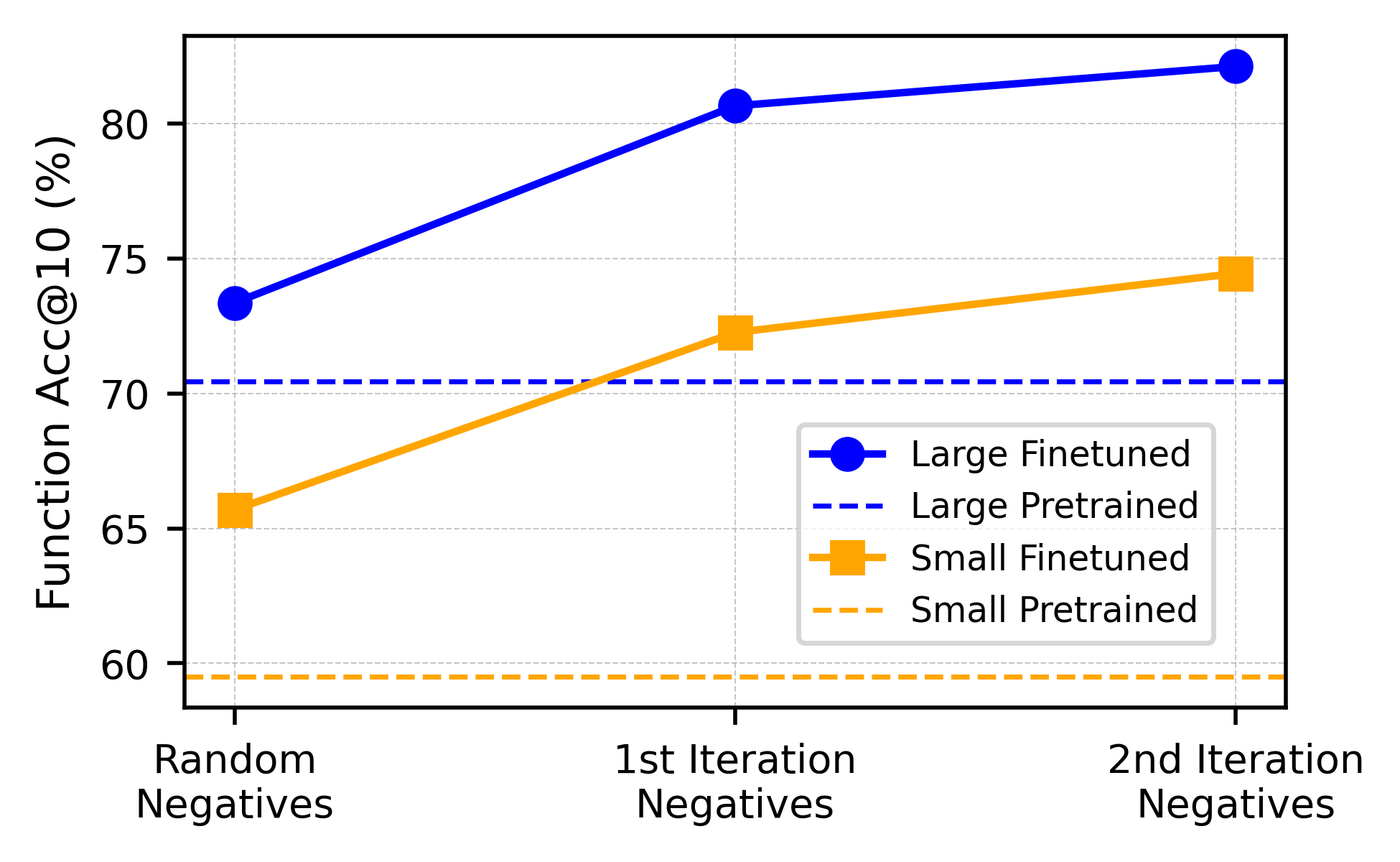}
\vspace{-1.8em}
\caption{Finetuned models notably improve from an additional iteration of negative mining.}
\label{fig:negatives}
\vspace{-1em}
\end{wrapfigure}

Here, we demonstrate the benefit of iterative negative mining by examining the impact of negative hardness on \textsc{SweRankEmbed} performance. Figure~\ref{fig:negatives} shows localization accuracy for Large and Small variants (finetuned and pretrained) with increasingly hard negatives. In an iterative mining approach, 1st iteration negatives are mined using the small pretrained model, and 2nd iteration negatives use the small model from 1st iteration. Results indicate that finetuning with random negatives yields smaller gains, while using 2nd iteration negatives yields notably improves performance over the 1st iteration.

\section{Multilingual Generalization}
\label{sec:multilingual}

Although \textsc{SweLoc} is constructed primarily from Python repositories, we hypothesize that \textsc{SweRank} generalizes effectively to other languages because the underlying base models (\textsc{CodeRankEmbed} and \textsc{GTE-Qwen2}) are pretrained on massive multilingual corpora.

To empirically validate this, we evaluate the models on SWE-Bench Multilingual~\citep{yang2025swesmith}, which includes 234 tasks across 9 languages (C, C++, Java, JavaScript, TypeScript, Rust, PHP, Ruby, and Go). We compare the performance against Gemini-Embedding~\citep{lee2025gemini}, a state-of-the-art proprietary general-purpose retriever.


\begin{table}[h]
    \centering
    \small
    \renewcommand{\arraystretch}{1.25}
    \setlength{\tabcolsep}{5pt}
    \begin{tabular}{@{}ll cc@{}}
        \toprule
        \textbf{Method} & \textbf{Model} & \textbf{Acc@5} & \textbf{Acc@10} \\
        \midrule
        \multirow{5}{*}{Retriever} 
          & CodeRankEmbed (137M) & 26.50 & 35.04 \\
          & SweRankEmbed-Small (137M) & 33.33 & 44.02 \\
          & GTE-Qwen2-7B-Instruct (7B) & 34.19 & 42.31 \\
          & SweRankEmbed-Large (7B) & \textbf{39.74} & \textbf{50.85} \\
          & \cellcolor{gray!10}Gemini-Embedding & \cellcolor{gray!10}36.75 & \cellcolor{gray!10}47.44 \\
        \midrule
        \multirow{3}{*}{Reranker} 
          & SweRankEmbed-Small (Base) & 33.33 & 44.02 \\
          & + CodeRankLLM & 42.74 & 51.28 \\
          & + SweRankLLM-Small & \textbf{49.15} & \textbf{56.84} \\
        \bottomrule
    \end{tabular}
    \caption{Function Localization Performance on SWE-Bench Multilingual. \textsc{SweRank} generalizes effectively to non-Python languages.}
    \label{tab:multilingual}
\end{table}

The results, summarized in Table~\ref{tab:multilingual}, support our hypothesis. \textsc{SweRankEmbed-Large} (50.85\% Acc@10) outperforms the proprietary Gemini-Embedding (47.44\%), despite being finetuned on Python data. Furthermore, finetuning on \textsc{SweLoc} provided significant gains over the base models for both the retriever and the reranker. This demonstrates that the ``issue-to-code'' relevance signal learned from \textsc{SweLoc} is not language-specific and transfers effectively across different languages.
\newpage
\section{Efficiency Comparison with Agentic Baseline}
\label{sec:latency_comparison_multiagent}
\subsection{Inference Latency Analysis}
\label{sec:latency_analysis}

\begin{wraptable}{r}{0.4\linewidth}
    \centering
    \small
    \vspace{-1.1em}
    \renewcommand{\arraystretch}{1.25}
    \setlength{\tabcolsep}{4pt}
    \resizebox{\linewidth}{!}{%
    \begin{tabular}{@{}ll c@{}}
        \toprule
        \textbf{Approach} & \textbf{Model} & \textbf{Latency (s)} \\
        \midrule
        SweRank & SweRankLLM (7B) & \textbf{12.5} \\
        SweRank & GPT-4o & 30.2 \\
        LocAgent & GPT-4o & 85.3 \\
        \bottomrule
    \end{tabular}%
    }
    \vspace{-1em}
    \caption{Inference Latency comparison.}
    \label{tab:latency_only}
    \vspace{-2em}
\end{wraptable}

To complement the cost analysis, we evaluate the inference latency of our approach compared to LocAgent. The average latency is measured over 50 instances on SWE-Bench-Lite. For the \textsc{SweRank} framework, the retrieval embeddings can be pre-computed and indexed, making the online retrieval cost negligible; therefore, the primary latency bottleneck stems purely from the reranking step.

Table~\ref{tab:latency_only} summarizes the results. When deploying the \textsc{SweRankLLM-Small} model locally on a single 80GB A100 GPU, the system achieves an average latency of just 12.5 seconds per instance. This is approximately \textbf{7$\times$ faster} than the LocAgent baseline, making \textsc{SweRank} far more viable for real-time developer assistance scenarios where rapid feedback is critical. Even when controlling for the underlying model by using GPT-4o for both approaches, \textsc{SweRank} remains nearly \textbf{3$\times$ faster}. This efficiency gain primarily comes from \textsc{SweRank} resolving the issue in a single-turn ranking pass, whereas agentic baselines like LocAgent relying on multi-turn loops, involving iterative thought generation, tool execution, and context reading, which naturally accumulates significant latency.

\subsection{Token Efficiency and Footprint}
\label{sec:token_analysis}

\begin{wraptable}{r}{0.4\linewidth}
    \centering
    \small
    \vspace{-1.1em}
    \renewcommand{\arraystretch}{1.25}
    \setlength{\tabcolsep}{4pt}
    \resizebox{\linewidth}{!}{%
    \begin{tabular}{@{}l cc@{}}
        \toprule
        \textbf{Approach} & \textbf{Prompt Tokens} & \textbf{Output Tokens} \\
        \midrule
        SweRank & 78,409 & 741 \\
        LocAgent & 234,197 & 1,884 \\
        \bottomrule
    \end{tabular}%
    }
    \vspace{-0.5em}
    \caption{Token footprint comparison.}
    \label{tab:tokens_only}
    \vspace{-1em}
\end{wraptable}

Beyond latency, the computational load of a system is heavily influenced by its token usage. Agent-based approaches often suffer from "context bloat," as they must maintain a running history of all past thoughts, observations, and tool outputs throughout the interaction loop. We analyzed the average token footprint (Average Input Prompt \& Output Tokens) required for each github issue.

As shown in Table~\ref{tab:tokens_only}, \textsc{SweRank} operates with a \textbf{$\sim$3$\times$ lower} input token footprint compared to the agentic baseline. By formulating localization as a ranking problem rather than a sequential decision-making process, \textsc{SweRank} eliminates the need for extensive history management. Furthermore, the reduction in output tokens is even more pronounced. Since output tokens are significantly more expensive and slower to generate than input tokens, this reduction directly translates to the lower latency observed in \S{\ref{sec:latency_analysis}} and substantially reduced inference costs. This confirms that \textsc{SweRank} provides a more sustainable and scalable alternative to agentic loops for issue localization.


\section{Qualitative Examples}
\label{sec:qualitative}
Figure ~\ref{fig:case_study} presents qualitative examples from SWE-Bench-Lite where \textsc{SweRank} correctly localizes the function to edit while LocAgent is unable to. In both instances, LocAgent incorrectly identifies functions that likely correspond to where the problem manifests rather than where it originates. 

\begin{figure}[!htb]
    \centering
    \includegraphics[width=\linewidth]{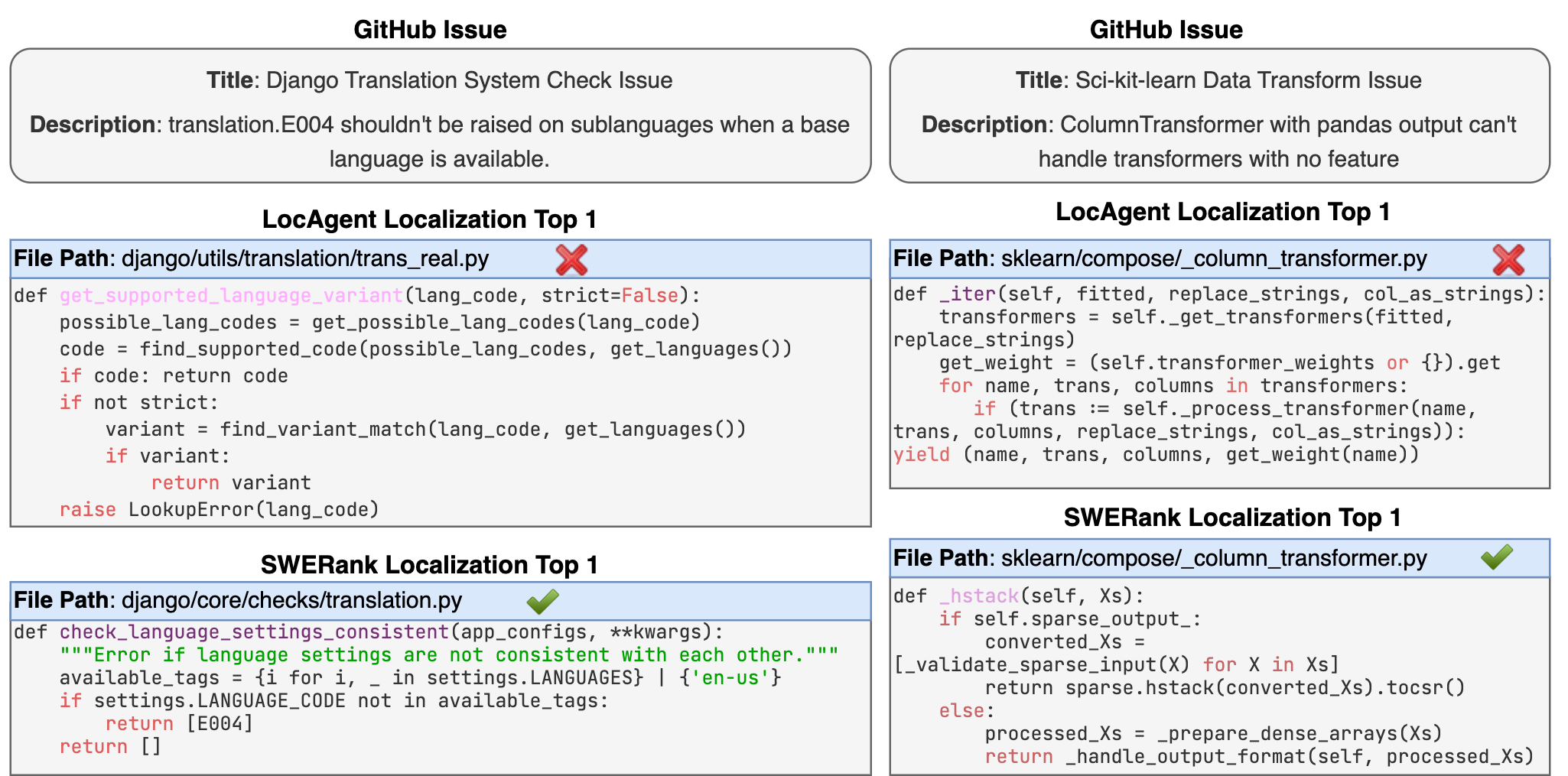}
    \caption{Examples from SWE-Bench-Lite where LocAgent mislocalizes the function,
while our SWERank framework does function localization correctly}
    \label{fig:case_study}
\end{figure}

\section{Diversity of Issue Topics in \textsc{SweLoc}}
\label{sec:sweloc_topics}
To provide more insight into the variety and complexity of issue topics in \textsc{SweLoc}, we analyze the distribution of topics for 10k randomly sampled instances. We use Nomic Atlas\footnote{\url{https://atlas.nomic.ai/}}, a popular unstructured text visualization tool, that employs a cluster-based keyword identification algorithm and leverages a language model to generate topics. Figure \ref{fig:topic_barplot} shows the frequency of top-15 topics.

\begin{figure}[!htb]
    \centering
    \includegraphics[width=0.9\linewidth]{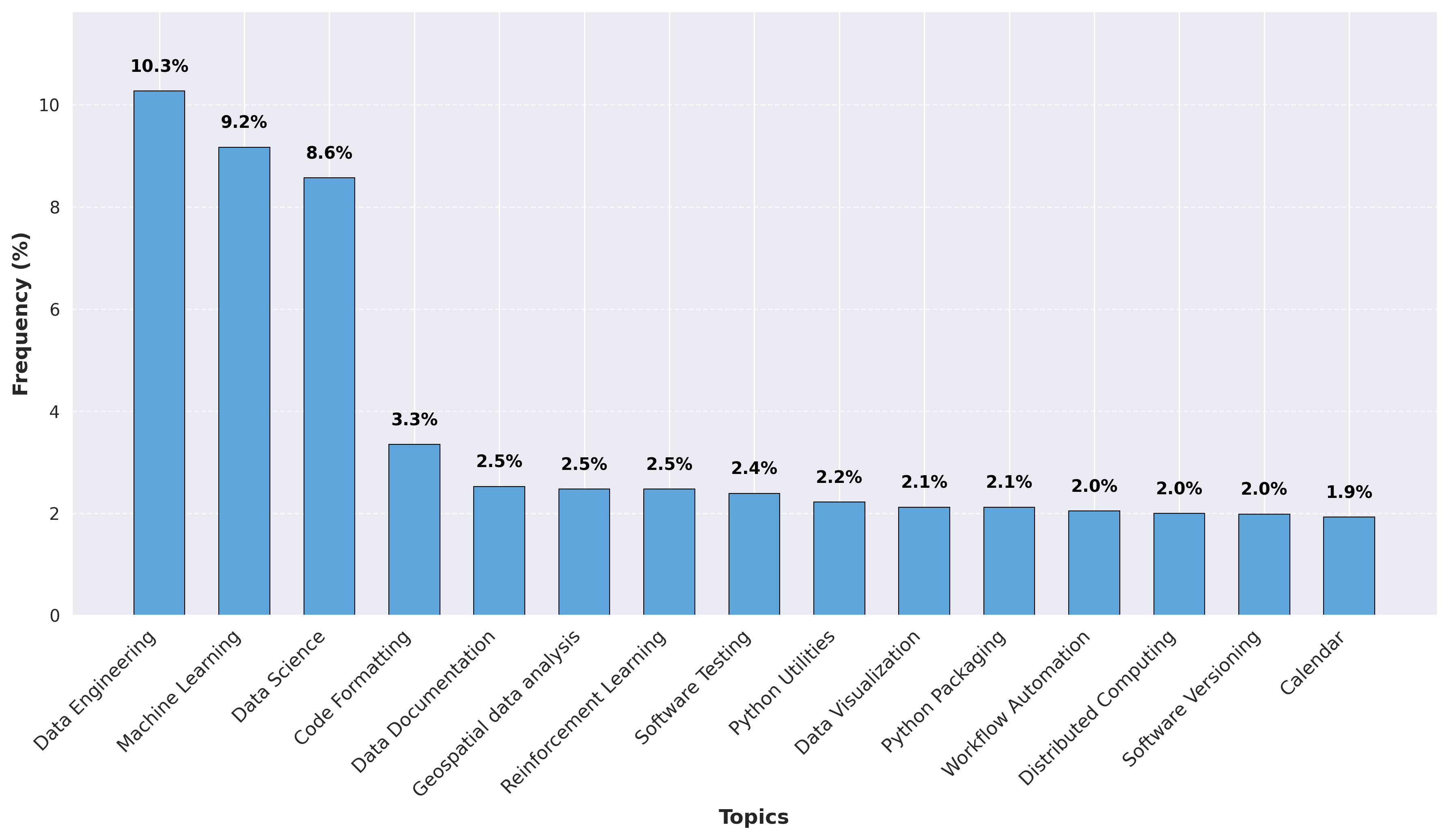}
    \vspace{-0.5em}
    \caption{Top-15 issue topics and their frequencies from a randomly sampled subset of \textsc{SweLoc}.}
    \label{fig:topic_barplot}
\end{figure}

\end{document}